\documentclass[preprint,prd,nofootinbib,tightenlines,groupedaddress,amsmath,amssymb]{revtex4}

\usepackage{bm}
\usepackage{color}
\usepackage{graphicx}
\usepackage{multirow}

\begin{document}
\baselineskip=15pt \parskip=3pt

\vspace*{3em}

\title{Seesaw Models with Minimal Flavor Violation}

\author{Xiao-Gang He,$^{1,2,3}$ Chao-Jung Lee$^{2}$, Jusak Tandean$^{2}$, Ya-Juan Zheng$^2$}
\affiliation{${}^{1}$INPAC, SKLPPC, and Department of Physics,
Shanghai Jiao Tong University, Shanghai 200240, China \vspace{3pt} \\
${}^{2}$CTS, CASTS, and Department of Physics, National Taiwan University,
Taipei 106, Taiwan \vspace{3pt} \\
${}^{3}$National Center for Theoretical Sciences and Physics Department of National
Tsing Hua University, Hsinchu 300, Taiwan \vspace{3ex}}


\begin{abstract}
We explore realizations of minimal flavor violation (MFV) for leptons in the simplest seesaw
models where the neutrino mass generation mechanism is driven by new fermion singlets (type I)
or triplets (type~III) and by a scalar triplet (type II).  We also discuss similarities and
differences of the MFV implementation among the three scenarios.
To study the phenomenological implications, we consider a~number of effective dimension-six
operators that are purely leptonic or couple leptons to the standard-model gauge and Higgs
bosons and evaluate constraints on the scale of MFV associated with these operators from
the latest experimental information.
Specifically, we employ the most recent measurements of neutrino mixing parameters as well as
the currently available data on flavor-violating radiative and three-body decays of charged
leptons, \,$\mu\to e$\, conversion in nuclei, the anomalous magnetic moments of charged
leptons, and their electric dipole moments.
The most stringent lower-limit on the MFV scale comes from the present experimental bound on
\,$\mu\to e\gamma$\, and can reach 500 TeV or higher, depending on the details of the seesaw
scheme.
With our numerical results, we illustrate some important differences among the seesaw types.
In particular, we show that in types I and III there are features which can bring about
potentially remarkable effects which do not occur in type II.
In addition, we comment on how one of the new effective operators can induce flavor-changing
dilepton decays of the Higgs boson, which may be probed in upcoming searches at the LHC.
\end{abstract}

\maketitle

\section{Introduction}

The standard model (SM) of particle physics has been immensely successful in describing
a~vast amount of experimental data at energies up to $\cal O$(100)\,GeV~\cite{pdg}.
One of the major implications is that in the quark sector flavor-dependent new interactions
beyond the SM can readily be ruled out if they give rise to substantial flavor-changing
neutral currents (FCNC).
This motivates the formulation of the principle of so-called minimal flavor violation (MFV),
which postulates that the sources of all FCNC and $CP$ violation reside in SM renormalizable
Yukawa couplings defined at tree level~\cite{mfv1,D'Ambrosio:2002ex}.
The MFV framework offers a predictive and systematic way to explore new physics
which does not conserve quark flavor and $CP$ symmetries.

The implementation of the MFV principle for quarks is straightforward, but for leptons
there are ambiguities, as the SM by itself does not predict lepton-flavor violation.
Since there is now compelling empirical evidence for neutrino masses and mixing~\cite{pdg},
it is of interest to formulate MFV in the lepton sector by incorporating ingredients beyond
the SM that can account for this observation~\cite{Cirigliano:2005ck}.
However, we lack knowledge regarding not only the origin of neutrino mass, but also
the precise nature of massive neutrinos.
They could be Dirac fermions, like the electron and quarks, as far as their spin
properties are concerned.
Alternatively, neutrinos being electrically neutral, there is also the possibility that they
are Majorana particles.
The mass-generation mechanisms and Yukawa couplings for the neutrinos in the two cases
differ significantly.
Since the MFV hypothesis is closely associated with Yukawa couplings, one expects that
the resulting phenomenologies in the two scenarios are also different.

In the Majorana neutrino case, there have been studies in the literature on some aspects of
MFV realizations in various seesaw scenarios~\cite{Cirigliano:2005ck,Cirigliano:2006su,
Branco:2006hz,Davidson:2006bd,Gavela:2009cd}, especially in the well-known simplest models
of types I, II, and\,\,III\,\,\cite{seesaw1,seesaw12,seesaw2,seesaw3}.
In this work, we take another look at these three seesaw schemes to investigate
the contributions of new interactions organized according to the MFV principle.
We adopt a model-independent approach, where such contributions consist of an infinite number
of terms which are built up from leptonic Yukawa couplings and their products.
It turns out that the infinite series can be resummed into only 17 terms~\cite{Colangelo:2008qp}.
This formulation allows one to have a much more compact understanding of the terms
that pertain to a given process.
We find that for the specific processes to be considered only a few of them are relevant.
We apply this to extract lower limits on the scale of MFV in the three seesaw scenarios using
the latest experimental data, including the existing bounds on flavor-violating leptonic
processes and the most recent measurements of neutrino mixing parameters.
Also, we will examine the similarities and differences among the three seesaw types in
relation to their MFV phenomenologies.
We demonstrate especially that in types I and III there are features which can bring about
potentially remarkable effects which do not happen in type II.

The plan of this paper is as follows.
In the next section, we describe the MFV framework for leptons in the case that neutrinos
are Dirac fermions.
In Section~\ref{mfvseesaw}, we discuss the implementation of the MFV principle in
scenarios involving Majorana neutrinos with masses generated via the seesaw mechanism
of types I, II, and~III.
This is applied in Section~\ref{dipole}, where we explore some of the phenomenology
with effective dipole operators involving leptons and the photon.
We evaluate constraints on the MFV scale associated with these operators from currently
available data on flavor-violating radiative decays of charged leptons, \,$\mu\to e$\,
conversion in nuclei, flavor-violating three-body decays of charged leptons, and their
anomalous magnetic moments and electric dipole moments.
With our numerical results, we illustrate some striking differences among the three
seesaw types.
In Section \ref{sec:ex-op}, we look at several other leptonic operators satisfying
the MFV principle.
One of them can cause flavor-violating decay of the Higgs boson, which is testable at the LHC.
We make our conclusions in Section~\ref{conclusion}.

\section{Leptonic MFV with Dirac neutrinos\label{mfvdirac}}

Let us begin by describing how to arrange interactions under the MFV framework for leptons
assuming that neutrinos are of Dirac nature.
Following the MFV hypothesis that renormalizable Yukawa couplings defined at tree level are
the only sources of FCNC and $CP$ violation, we need to start with such couplings for
the neutrinos and charged leptons.
We slightly extend the SM by including three right-handed neutrinos which transform
as $(1,1,0)$ under the SM gauge group
\,${\cal G}_{\scriptscriptstyle\rm SM}={\rm SU}(3)_C\times$SU(2)$_L\times$U(1)$_Y$.\,
The Lagrangian responsible for the lepton masses can then be written as
\begin{eqnarray} \label{lepton-YC}
{\cal L}_{\rm m}^{} \,\,=\,\, -(Y_\nu)_{kl}^{}\,\overline{L}_{k,L\,}^{}\nu_{l,R\,}^{}\tilde H
\,-\, (Y_e)_{kl}^{}\,\overline{L}_{k,L\,}^{}E_{l,R\,}^{}H  \;+\; {\rm H.c.} \,,
\end{eqnarray}
where summation over \,$k,l=1,2,3$\, is implicit, $Y_{\nu,e}$ are Yukawa coupling matrices,
$L_{k,L}$ represents left-handed lepton doublets, $\nu_{k,R}$ and
$E_{k,R}$ denote right-handed neutrinos and charged leptons, $H$ is the Higgs boson doublet,
and \,$\tilde H=i\tau_2^{}H^*$\, involving the second Pauli matrix~$\tau_2^{}$.
Under the SM gauge group, $L_{k,L}$, $E_{k,R}$, and $H$ transform as
\,$(1, 2, -1/2)$, \,$(1,1, -1)$,\, and \,$(1, 2, 1/2)$,\, respectively.

The MFV hypothesis\,\,\cite{D'Ambrosio:2002ex,Cirigliano:2005ck} implies that ${\cal L}_{\rm m}$
is formally invariant under the global group
\,U(3)$_L\times$U(3)$_\nu\times$U(3)$_E=G_\ell\times$U(1)$_L\times$U(1)$_\nu\times$U(1)$_E$\,
with \,$G_\ell=\rm SU$(3)$_L\times$SU(3)$_\nu\times$SU(3)$_E$.\,
This entails that $L_{k,L}$, $\nu_{k,R}$, and $E_{k,R}$ belong to the
fundamental representations of the SU(3)$_{L,\nu,E}$, respectively,
\begin{eqnarray}
L_L^{} \,\to\, V_L^{}L_L^{} \,, ~~~~~~~ \nu_R^{} \,\to\, V_\nu^{}\nu_R^{} \,, ~~~~~~~
E_R^{} \,\to\, V_E^{}E_R^{} \,, ~~~~~~~ V_{L,\nu,E} \,\in\, {\rm SU}(3)_{L,\nu,E} \,,
\end{eqnarray}
and under $G_\ell$ the Yukawa couplings transform in the spurion sense according to
\begin{eqnarray}
Y_\nu^{} \,\to\, V_L^{}Y_\nu^{}V^\dagger_\nu \,\,\sim\,\, (3,\bar 3,1) \,,
~~~~~~~ Y_e^{} \,\to\, V_L^{}Y_e^{}V^\dagger_E \,\,\sim\,\, (3,1,\bar 3) \,.
\end{eqnarray}

Taking advantage of the requirement that the final effective Lagrangian be invariant
under~$G_\ell$, without loss of generality one can always work in the basis where $Y_e$ is
diagonal,
\begin{eqnarray} \label{Ye}
Y_e \,\,=\,\, \frac{\sqrt2}{v}\, {\rm diag}\bigl(m_e^{},m_\mu^{},m_\tau^{}\bigr)
\end{eqnarray}
with \,$v\simeq246$\,GeV\, being the Higgs's vacuum expectation value (VEV),
and $\nu_{k,L}$, $\nu_{k,R}$, $E_{k,L}$, and $E_{k,R}$ refer to the mass eigenstates.
Consequently, one can express $L_{k,L}$ and $Y_\nu$ in terms of
the Pontecorvo-Maki-Nakagawa-Sakata
neutrino mixing matrix $U_{\scriptscriptstyle\rm PMNS}$~as
\begin{eqnarray}
L_{k,L}^{} \,= \left( \!\begin{array}{c} (U_{\scriptscriptstyle\rm PMNS})_{kl\,}^{}
\nu_{l,L}^{} \vspace{2pt} \\ E_{k,L}^{} \end{array}\! \right) , ~~~~~~~
Y_\nu \,=\, \frac{\sqrt2}{v}\,U_{\scriptscriptstyle\rm PMNS}^{}\,\hat m_\nu^{} \,, ~~~~
\hat m_\nu^{} \,=\, {\rm diag}\bigl(m_1^{},m_2^{},m_3^{}\bigr) \,, \label{YnuD}
\end{eqnarray}
where $m_{1,2,3}^{}$ are the light neutrino eigenmasses and in the standard
parametrization~\cite{pdg}
\begin{eqnarray} \label{pmns}
U_{\scriptscriptstyle\rm PMNS}^{} \,= \left(\!\begin{array}{ccc}
 c_{12\,}^{}c_{13}^{} & s_{12\,}^{}c_{13}^{} & s_{13}^{}\,e^{-i\delta}
\vspace{1pt} \\
-s_{12\,}^{}c_{23}^{}-c_{12\,}^{}s_{23\,}^{}s_{13}^{}\,e^{i\delta} & ~~
 c_{12\,}^{}c_{23}^{}-s_{12\,}^{}s_{23\,}^{}s_{13}^{}\,e^{i\delta} ~~ & s_{23\,}^{}c_{13}^{}
\vspace{1pt} \\
 s_{12\,}^{}s_{23}^{}-c_{12\,}^{}c_{23\,}^{}s_{13}^{}\,e^{i\delta} &
-c_{12\,}^{}s_{23}^{}-s_{12\,}^{}c_{23\,}^{}s_{13}^{}\,e^{i\delta} & c_{23\,}^{}c_{13}^{}
\end{array}\right) ,
\end{eqnarray}
with $\delta$ being the Dirac $CP$-violation phase, \,$c_{kl}^{}=\cos\theta_{kl}^{}$,\, and
\,$s_{kl}^{}=\sin\theta_{kl}^{}$.\,

Based on the transformation properties of the fields and Yukawa spurions, one then uses
an arbitrary number of Yukawa coupling matrices to put together $G_\ell^{}$-invariant objects
which can induce the desired FCNC and $CP$-violating interactions.
Thus, for operators involving two lepton fields, the pertinent building blocks are
\begin{eqnarray}
\overline{L}_L^{}\gamma_\alpha^{}\Delta_l^{}L_L^{} \;, ~~~~
\overline{\nu}_R^{}\gamma_\alpha^{}\Delta_{\nu8}\nu_R \;, ~~~~
\overline{E}_R^{}\gamma_\alpha^{}\Delta_{e8}^{}E_R^{} \;, ~~~~
\overline{\nu}_R^{}\bigl(1,\sigma_{\alpha\beta}^{}\bigr)\Delta_\nu^{}L_L^{} \;, ~~~~
\overline{E}_R^{}\bigl(1,\sigma_{\alpha\beta}^{}\bigr)\Delta_e^{}L_L^{} \;.
\end{eqnarray}
For these to be $G_\ell$ invariant, the $\Delta$'s should transform according to
\begin{eqnarray} \label{Deltai}
\Delta_l^{} &\sim& (1\oplus8,1,1)\;, ~~~~ \Delta_{\nu8}^{}\,\sim\,(1,1\oplus8,1)\;, ~~~~
\Delta_{e8}^{}\,\sim\,(1,1,1\oplus8)\;,
\nonumber \\
\Delta_\nu^{} &\sim& (\bar 3,3,1)\;, \hspace{8ex} \Delta_e^{}\,\sim\,(\bar 3,1,3) \;.
\end{eqnarray}
Since \,$\overline{L}_L^{}\gamma_\alpha^{}\Delta_l^{}L_L^{}$,
\,$\overline{\nu}_R^{}\gamma_\alpha^{}\Delta_{\nu8}^{}\nu_R^{}$,\,
and \,$\overline{E}_R^{}\gamma_\alpha^{}\Delta_{e8}^{}E_R^{}$\,  must be Hermitian,
$\Delta_{l,\nu8,e8}^{}$ must be Hermitian as well.
To be acceptable terms in the Lagrangian, the above objects should be combined with
appropriate numbers of other SM fields into singlets under the SM gauge group,
with all the Lorentz indices contracted.

The MFV principle dictates that these $\Delta$'s are built up from the Yukawa coupling matrices
$Y_{\nu,e}^{}$ and~$Y_{\nu,e}^\dagger$.
Let us first discuss a nontrivial $\Delta$ which transforms as $(1\oplus8,1,1)$ under $G_\ell$
and consists of terms in powers of
\begin{eqnarray} \label{ABD}
{\sf A} \,\,=\,\, Y_\nu^{}Y^\dagger_\nu \,\,=\,\,
\frac{2}{v^2}\, U_{\scriptscriptstyle\rm PMNS\,}^{} \hat m_\nu^2
U_{\scriptscriptstyle\rm PMNS}^\dagger \;, ~~~~~~~
{\sf B} \,\,=\,\, Y_e^{}Y_e^\dagger \,\,=\,\,
\frac{2}{v^2}\, {\rm diag}\bigl(m_e^2,m_\mu^2,m_\tau^2\bigr) \,,
\end{eqnarray}
both of which also transform as $(1\oplus8,1,1)$.
Formally, $\Delta$ is a sum of infinitely many terms,
\,$\Delta=\sum\xi_{jkl\cdots\,}^{}{\sf A}^j{\sf B}^k{\sf A}^l\cdots$\, with coefficients
$\xi_{jkl\cdots}^{}$ expected to be at most of~${\cal O}$(1).
Under the MFV framework, these coefficients must be real because otherwise they would
introduce new sources of $CP$ violation beyond those in the Yukawa couplings.
With the Cayley-Hamilton identity
\,$X^3=X^2\,{\rm Tr}X+\mbox{$\frac{1}{2}$}_{\,}X\bigl[{\rm Tr}X^2-({\rm Tr}X)^2\bigr]
+\openone{\rm Det}X$\,
for an invertible 3$\times$3 matrix~$X$, one can resum the infinite series into a limited
number of terms~\cite{Colangelo:2008qp},
\begin{eqnarray}
\Delta &\,=\,& \xi^{}_1\openone + \xi^{}_{2\,}{\sf A}+\xi^{}_{3\,}{\sf B}
+ \xi^{}_{4\,}{\sf A}^2 +\xi^{}_{5\,}{\sf B}^2 + \xi^{}_{6\,}{\sf AB} + \xi^{}_{7\,}{\sf BA}
+ \xi^{}_{8\,}{\sf ABA} + \xi^{}_{9\,}{\sf BA}^2 + \xi^{}_{10\,}{\sf BAB}
\nonumber \\ && \! +~
\xi^{}_{11\,}{\sf AB}^2 + \xi^{}_{12\,}{\sf ABA}^2 + \xi^{}_{13\,}{\sf A}^2{\sf B}^2
+ \xi^{}_{14\,}{\sf B}^2{\sf A}^2 + \xi^{}_{15\,}{\sf B}^2{\sf AB}
+ \xi^{}_{16\,}{\sf AB}^2{\sf A}^2+\xi^{}_{17\,}{\sf B}^2{\sf A}^2{\sf B} ~, ~~~ \label{general}
\end{eqnarray}
where $\openone$ stands for the 3$\times$3 unit matrix.
Though one starts with all $\xi_{jkl\cdots\,}^{}$ being real, the resummation process generally
renders the coefficients $\xi_r^{}$ in Eq.\,(\ref{general}) complex due to imaginary
parts created among the traces of the matrix products \,${\sf A}^j{\sf B}^k{\sf A}^l\cdots$\,
with \,$j+k+l+\cdots\ge6$\, after the application of the Cayley-Hamilton identity.
The imaginary contributions turn out to be reducible to factors proportional to a Jarlskog
invariant quantity,
\,${\rm Im\,Tr}\bigl({\sf A}^2{\sf BAB}^2\bigr)=(i/2)\,{\rm Det}[{\sf A,B}]$,\,
which is much smaller than unity~\cite{Colangelo:2008qp}.

With Eq.\,(\ref{general}), one can devise the objects in Eq.\,(\ref{Deltai}).
Thus, the first of the Hermitian combinations can be \,$\Delta_l=\Delta+\Delta^\dagger$.\,
To obtain nontrivial~$\Delta_{\nu,e}$, one can take \,$\Delta_\nu^{}=Y_\nu^\dagger\Delta$\,
and \,$\Delta_e^{}=Y_e^\dagger\Delta$.\,
The construction of $\Delta_{\nu8,e8}$ can be carried out in a similar way, except $\sf A$ and
$\sf B$ are replaced by \,$\tilde A=Y_\nu^\dagger Y_\nu^{}$\, and
\,$\tilde B=Y_e^\dagger Y_e^{}$.\,
Since $\tilde A$ and $\tilde B$ are diagonal, so are any powers of them.
Therefore, $\Delta_{\nu8,e8}$ do not produce any FCNC and $CP$-violation effects.

We end this section by mentioning that the above discussion can be easily applied to
the quark sector with the renormalizable Yukawa Lagrangian given by
\begin{eqnarray}
{\cal L}_{\rm m}^{} &\,=\,& -(Y_u)_{kl}^{}\,\overline{Q}_{k,L\,}^{}U_{l,R\,}^{} \tilde H
- (Y_d)_{kl}^{}\,\overline{Q}_{k,L\,}^{}D_{l,R\,}^{} H \;+\; {\rm H.c.} \,,
\label{quark-YC}
\end{eqnarray}
where $Y_{u,d}$ are Yukawa coupling matrices, $Q_{k,L}$ represents left-handed quark doublets,
and $U_{k,R}$ $(D_{k,R})$ denote right-handed up-type (down-type) quarks.
These fields transform as \,$(3, 2, 1/6)$, \,$(3, 1, 2/3)$,\,  and \,$(3,1, -1/3)$,\,
respectively, under the SM gauge group ${\cal G}_{\scriptscriptstyle\rm SM}$.
In the basis where $Y_d$  is diagonalized,
\begin{eqnarray}
Y_d^{} \,\,=\,\, \frac{\sqrt2}{v}\, {\rm diag}\bigl(m_d^{},m_s^{},m_b^{}\bigr) \,, ~~~~~~~
Y_u^{} \,\,=\,\, \frac{\sqrt2}{v}\,V^\dagger_{\scriptscriptstyle\rm {CKM}^{}}\,\hat m_u^{} \,,
~~~~ \hat m_u^{} \,\,=\,\, {\rm diag}\bigl(m_u^{},m_c^{},m_t^{}\bigr) \,, \label{Yq}
\end{eqnarray}
where $V_{\scriptscriptstyle\rm CKM}$ is the Cabibbo-Kobayashi-Maskawa matrix which has
the same standard parametrization as in Eq.\,(\ref{pmns}).
For MFV interactions, employing $Y_{u,d}$ along with \,${\sf A}=Y_u^{}Y^\dagger_u$\, and
\,${\sf B}=Y_d^{}Y^\dagger_d$\, as building blocks, one can construct objects such as
$\Delta_q$, $\Delta_u$, and $\Delta_d$, which are the quark counterparts of $\Delta_l$,
$\Delta_\nu$, and~$\Delta_e$, respectively \cite{He:2014fva}.

\section{Seesaw models with MFV\label{mfvseesaw}}

If neutrinos are Majorana particles, the Yukawa couplings that take part in generating
their masses differ from those in the Dirac neutrino case and depend on the model details.
In this section we discuss how to realize the MFV hypothesis in the well-motivated seesaw models.
The seesaw mechanism endows neutrinos with Majorana mass and provides a natural explanation for
why they are much lighter than their charged partners.
If just one kind of new particle is added to the minimal SM, there are three different
scenarios \cite{seesaw1,seesaw12,seesaw2,seesaw3}:
the famous seesaw models of type I, type~II, and type III.
A crucial step in the implementation of MFV in a given model is to identify the quantities
$\sf A$ and $\sf B$ in terms of the relevant Yukawa couplings.
This will be the emphasis of the section.

\subsection{MFV in type-I seesaw model}

In the type-I seesaw model, the SM is slightly expanded with the inclusion of three
gauge-singlet right-handed neutrinos, $\nu_{k,R}^{}$, which each carry a lepton number of 1
and are allowed to possess Majorana masses~\cite{seesaw1}.
The renormalizable Lagrangian for the lepton masses is
\begin{eqnarray} \label{majorana-neu}
{\cal L}_{\rm m}^{\rm I} \,\,=\,\, -(Y_\nu)_{kl}^{}\,\overline{L}_{k,L\,}^{}\nu_{l,R\,}^{}\tilde H
- (Y_e)_{kl}^{}\,\overline{L}_{k,L\,}^{}E_{l,R\,}^{} H
- \mbox{$\frac{1}{2}$}\, (M_\nu)_{kl}^{}\,\overline{\nu^{\rm c}_{k,R}}\,\nu_{l,R}^{}
\;+\; {\rm H.c.} \,,
\end{eqnarray}
where \,$M_\nu={\rm diag}(M_1,M_2,M_3)$\, contains the right-handed neutrinos' Majorana masses,
breaking lepton-number symmetry, and \,$\nu^{\rm c}_{k,R}\equiv(\nu_{k,R})^{\rm c}$,\,
the superscript c referring to charge conjugation.
Accordingly, the masses of the neutral fermions make up the 6$\times$6 matrix
\begin{eqnarray}
{\sf M} \,\,=\,\, \left ( \begin{array}{cc} 0 & M_{\rm D}^{} \vspace{2pt} \\
M_{\rm D}^{\rm T} & M_\nu^{} \end{array} \right)
\end{eqnarray}
in the $\bigl(U_{\scriptscriptstyle\rm PMNS}^*(\nu_L)^{\rm c}, \nu_R^{}\bigr){}^{\rm T}$ basis,
where \,$M_{\rm D}=v Y_\nu/\sqrt2$.\,
If the nonzero elements of $M_\nu$ are much greater than those of $M_{\rm D}$,
the seesaw mechanism becomes operational~\cite{seesaw1}, resulting in the light neutrinos'
mass matrix
\begin{eqnarray} \label{mnu}
m_\nu \,\,=\,\, -\frac{v^2}{2}\, Y_\nu^{}M_\nu^{-1}Y_\nu^{\rm T} \,\,=\,\,
U_{\scriptscriptstyle\rm PMNS\,}^{}\hat m_{\nu\,}^{}U_{\scriptscriptstyle\rm PMNS}^{\rm T} \;,
\end{eqnarray}
where now $U_{\scriptscriptstyle\rm PMNS}$ contains the diagonal matrix
\,$P={\rm diag}(e^{i\alpha_1/2},e^{i\alpha_2/2},1)$\, multiplied from the right,
with $\alpha_{1,2}^{}$ being the $CP$-violating Majorana phases.
It follows that $Y_\nu$ in Eq.\,(\ref{YnuD}) is no longer valid,
and one can instead pick $Y_\nu$ to be~\cite{Casas:2001sr}
\begin{eqnarray} \label{YnuM}
Y_\nu^{} \,\,=\,\,
\frac{i\sqrt2}{v}\,U_{\scriptscriptstyle\rm PMNS\,}^{}\hat m^{1/2}_\nu O M_\nu^{1/2} \,,
\end{eqnarray}
where $O$ is a matrix satisfying \,$OO^{\rm T}=\openone$.\,
Since $O$ can be complex, it is a potentially important new source of $CP$ violation
besides $U_{\scriptscriptstyle\rm PMNS}$.

Another implication of the presence of $M_\nu$ is that it will introduce an additional
source of flavor violation if the masses $M_{1,2,3}$ of the right-handed neutrinos,
$\nu_{k,R}^{}$, are unequal\,\,\cite{Davidson:2006bd}.
In that case, the extra flavor spurions should generally be taken into account by adding their
contributions to $\Delta$ in\,\,Eq.\,(\ref{general}).
However, if $\nu_{k,R}^{}$ have the same mass, \,$M_{1,2,3}={\cal M}$,\, no new terms will
need to be included in Eq.\,(\ref{general}).
In our treatment of the type-I seesaw with MFV, we restrict ourselves to this least complicated
possibility that the right-handed neutrinos are degenerate.
It follows that in this instance the flavor symmetry is\,\,\cite{Cirigliano:2005ck}
\,$G_\ell=\rm SU$(3)$_L\times$O(3)$_\nu\times$SU(3)$_E$,\, under which
\,$\nu_R^{}\to{\cal O}_\nu^{}\nu_R^{}$\,
and \,$Y_\nu^{}\to V_L^{}Y_\nu^{}{\cal O}_\nu^{\rm T}$,\,
where ${\cal O}_\nu$ is an orthogonal real matrix.

Moreover, from Eq.\,(\ref{YnuM}) with \,$M_\nu={\cal M}\openone$\, and
$O$ being real or complex\,\,\cite{Cirigliano:2005ck,Branco:2006hz}
\begin{eqnarray} \label{AI}
{\sf A} \,\,=\,\, Y_\nu^{}Y^\dagger_\nu \,\,=\,\,
\frac{2\cal M}{v^2}\, U_{\scriptscriptstyle\rm PMNS\,}^{} \hat m^{1/2}_\nu O
O^\dagger\hat m^{1/2}_\nu U_{\scriptscriptstyle\rm PMNS}^\dagger \;,
\end{eqnarray}
whereas \,${\sf B}=Y_e^{}Y_e^\dagger$\, in Eq.\,(\ref{ABD}) is unchanged.
These matrices are to be incorporated into the MFV objects mentioned earlier,
such as $\Delta_{l,\nu,e}$.
We note that, since
\,$\tilde A=Y^\dagger_\nu Y_\nu^{}=(2{\cal M}/v^2)O^\dagger\hat m_\nu^{}O$,\,
here $\Delta_{\nu8}$ is no longer diagonal, but it pertains only to $\nu_R^{}$ interactions.

Now, in the Dirac neutrino case, $\sf A$ is given by Eq.\,(\ref{ABD}) and therefore
has tiny elements.
In contrast, if $\cal M$ is sufficiently large, the eigenvalues of
$\sf A$ in Eq.\,(\ref{AI}) can be sizable.
Nevertheless, since as an infinite series $\Delta$ has to converge, $\cal M$ cannot
be arbitrarily large~\cite{Colangelo:2008qp,He:2014fva}.
Accordingly, in numerical work we require the biggest eigenvalue of $\sf A$ to be unity,
which implies that the elements of $\sf B$\, are comparatively much smaller.

\subsection{MFV in type-II seesaw model}

The type-II seesaw model extends the SM with the addition of only one scalar SU(2)$_L$
triplet given by~\cite{seesaw12,seesaw2}
\begin{eqnarray}
T \,\,=\,\, \left(\begin{array}{cc} T^+/\sqrt2 & T^{++} \vspace{1pt} \\
T^{0} & -T^{+}/\sqrt2 \end{array}\right) \label{standpara}
\end{eqnarray}
which transforms as $(1, 3, 1)$ under the SM gauge group ${\cal G}_{\scriptscriptstyle\rm SM}$.
Accordingly, the Lagrangian describing the Yukawa couplings of leptons is
\begin{eqnarray} \label{LmII}
{\cal L}_{\rm m}^{\rm II} \,\,=\,\, -(Y_e)_{kl}^{}\,\overline{L}_{k,L\,}^{}E_{l,R\,}^{} H \,-\,
\mbox{$\frac{1}{2}$}\,(Y_T)_{kl}^{}\,\overline{L}_{k,L\,}^{}\tilde T\,
\big(L_{l,L}^{}\big)^{\rm c}\;+\;{\rm H.c.} \,,
\end{eqnarray}
with \,$\tilde T=i\tau_2^{}T^*$.\,
It respects lepton-number conservation if $T$ is assigned a lepton number of $-2$.
After the VEV of the neutral component of $T$ becomes nonzero,
\,$\bigl\langle T^0\bigr\rangle=v_T^{}/\sqrt2$,\, one obtains in ${\cal L}_{\rm m}^{\rm II}$
the neutrino mass matrix
\begin{eqnarray} \label{YT}
m_\nu^{} \,\,=\,\, \mbox{$\frac{1}{\sqrt2}$}\, v_{T\,}^{} Y_T^{} \,\,=\,\,
U_{\scriptscriptstyle\rm PMNS\,}^{}\hat m_{\nu\,}^{}U_{\scriptscriptstyle\rm PMNS}^{\rm T}
\end{eqnarray}
in the basis where the charged lepton's Yukawa coupling matrix, $Y_e$, has been diagonalized.
If the nonzero elements of $Y_T$ are of $\cal O$(1), the tiny size of neutrino masses
then comes from the suppression of $v_T^{}$ due to certain choices of the parameters in
the scalar potential $\cal V$.
One can express it as~\cite{Gelmini:1980re}
\begin{eqnarray}
{\cal V} &\,=\,&
-M_{H\,}^2 H^\dagger H + M_{T\,}^2{\rm Tr}(T^\dagger T)
+ \mbox{$\frac{1}{2}$}\lambda_{1\,}^{}(H^\dagger H)^2
+ \mbox{$\frac{1}{2}$}\lambda_{2\,}^{}[{\rm Tr}(T^\dagger T)]^2
+ \lambda_{3\,}^{}H^\dagger H\,{\rm Tr}(T^\dagger T)
\nonumber \\ && \! +\;
\lambda_{4\,}^{}{\rm Det}(T^\dagger T) + \lambda_{5\,}^{}H^\dagger T^\dagger T H
\,-\, \mbox{$\frac{1}{\sqrt2}$}\bigl( \mu_T^{}\, \tilde H^\dagger T^\dagger H \,+\,
\mu_T^*\, H^\dagger T\tilde H \bigr) \,,
\end{eqnarray}
where \,$M_H^2>0$\, and \,$M_T^2>0$.\,
The $\mu_T^{}$ part explicitly breaks lepton-number symmetry and causes $T^0$ to develop
a nonzero VEV.
From the minimization of $\cal V$, one gets
\begin{eqnarray} \label{vvT}
\lambda_1^{}v^2 \,\,=\,\, 2 M_H^2+2 |\mu_T^{}| v_T^{}-\lambda_3^{}v_T^2 \;, ~~~~~~~
|\mu_T^{}| v^2 \,\,=\,\, 2 M_{T\,}^2v_T^{}+\lambda_3^{}v^2 v_T^{}+\lambda_{2\,}^{}v_T^3 \;.
\end{eqnarray}
For \,$|\mu_T^{}|v_T^{}\ll M_H^2$\, and \,$v_T^{}\ll v$,\, the first equality simplifies to
\,$\lambda_1^{}v^2\simeq2 M_H^2$\, like in the SM, and with the additional conditions
\,$v\ll M_T^{}/|\lambda_3|^{1/2}$ and \,$v_T^{}\ll|\mu_T^{}|$\,
the second relation in Eq.\,(\ref{vvT}) translates into
\begin{eqnarray}
v_T^{} \,\,\simeq\,\, \frac{|\mu_T^{}|v^2}{2M_T^2} \;, \label{vTMT}
\end{eqnarray}
which is small if \,$|\mu_T^{}|v\ll M_T^2$.\,
This turns into the seesaw form \,$v_T^{}\sim v^2/M_T^{}$\, if \,$|\mu_T^{}|\sim M_T^{}\gg v$,\,
but the prerequisites just mentioned do not preclude a scenario with a relatively
lighter triplet, in which $M_T$ can be as low as the TeV level~\cite{Gavela:2009cd},
provided that \,$|\mu_T^{}|\ll v$.\,

Since the triplet couples to SM gauge bosons, a nonzero $v_T^{}$ will make the $\rho_0^{}$
parameter deviate from unity~\cite{Gelmini:1980re},
\begin{eqnarray}
\rho_0^{} \,\,=\,\, \frac{m_W^2}{m_{Z\,}^2\cos^{2\!}\theta_{\rm W}^{}} \,\,=\,\,
\frac{v^2+2 v_T^2}{v^2+4 v_T^2} \,\,\simeq\,\, 1 \,-\, \frac{2 v_T^2}{v^2} \;.
\end{eqnarray}
Its empirical value \,$\rho_0^{}=1.00040\pm0.00024$~\cite{pdg} then implies, at
the 2$\sigma$ level, that \,$v_T^{}<1.6$\,GeV.\,
This is much weaker than the requirement for the $v_T^{}$ range that can produce neutrino masses of
$\cal O$(0.1$\;$eV) if the elements of $Y_T$ are of $\cal O$(1).

To implement the MFV hypothesis in this seesaw scheme, one observes that the Lagrangian in
Eq.\,(\ref{LmII}) possesses formal invariance under the global group
\,U(3)$_L\times$U(3)$_E=G_\ell'\times$U(1)$_L\times$U(1)$_E$,\,
with \,$G_\ell'=\rm SU$(3)$_L\times$SU(3)$_E$,\, if
$L_L$ and $E_R$ belong to the fundamental representation of $SU(3)_{L,E}^{}$, respectively,
\begin{eqnarray}
L_L^{} \,\to\, V_L^{}L_L^{} \,, ~~~~~~~ E_R^{} \,\to\, V_E^{}E_R^{} \,, ~~~~~~~
V_{L,E} \,\in\, {\rm SU}(3)_{L,E} \,,
\end{eqnarray}
and the Yukawa couplings are spurions transforming according to
\begin{eqnarray}
Y_e^{} \,\to\, V_L^{}Y_e^{}V^\dagger_E \,, ~~~~~~~ Y_T^{} \,\to\, V_L^{}Y_T^{}V_L^{\rm T} \,.
\end{eqnarray}
Here the building block $\Delta$ still has the expression in Eq.\,(\ref{general}), with
\,${\sf B}=Y_e^{}Y_e^\dagger$\, being the same as in Eq.\,(\ref{ABD}), but unlike before
\begin{eqnarray} \label{AII}
{\sf A} \,\,=\,\, Y_T^{}Y_T^\dagger \,\,=\,\, \frac{2}{v_T^2}\,
U_{\scriptscriptstyle\rm PMNS\,}^{}\hat m_{\nu\,}^2 U_{\scriptscriptstyle\rm PMNS}^\dagger \,,
\end{eqnarray}
where from Eq.\,(\ref{YT})
\begin{eqnarray}
Y_T^{} \,\,=\,\, \frac{\sqrt2}{v_T^{}}\,
U_{\scriptscriptstyle\rm PMNS\,}^{}\hat m_{\nu\,}^{}U_{\scriptscriptstyle\rm PMNS}^{\rm T} \,.
\end{eqnarray}

It is interesting to notice that $\sf A$ in Eq.\,(\ref{AII}) is the same as its Dirac-neutrino
counterpart in~Eq.\,(\ref{ABD}), up to an overall factor.
Due to this difference, whereas the elements of the latter are tiny, those in
Eq.\,(\ref{AII}) can be of ${\cal O}(1)$ if $v_T^{}$ is similar in order of magnitude
to the neutrino masses.
This will in fact be realized in our numerical analysis, as we will again choose the largest
eigenvalue of $\sf A$ to be unity, which amounts to imposing
\,$v_T^{}=\sqrt2\;{\rm max}\bigl(m_1^{},m_2^{},m_3^{}\bigr)$.\,

Compared to Eq.\,(\ref{AI}) in the type-I case, $\sf A$ in Eq.\,(\ref{AII}) is in general
much simpler.
In particular, it no longer depends on the Majorana phases in $U_{\scriptscriptstyle\rm PMNS}$
which have canceled out due to $\hat m_\nu^{}$ being diagonal and does not involve the $O$
matrix which can supply potentially major extra effects including $CP$-violating
ones \cite{He:2014fva}.

\subsection{MFV in type-III seesaw model}

In the type-III seesaw model, the new particles consist only of three fermionic SU(2)$_L$
triplets~\cite{seesaw3}
\begin{eqnarray}
\Sigma_k^{} \,\,=\, \left(\!\begin{array}{cc} \Sigma_k^0/\sqrt2 & \Sigma_k^+ \vspace{2pt} \\
\Sigma_k^- & -\Sigma_k^0/\sqrt2 \end{array}\!\right) , ~~~~~~~
k \,\,=\,\, 1,2,3 \,,
\end{eqnarray}
which transform as $(1, 3, 0)$ under the SM gauge group ${\cal G}_{\scriptscriptstyle\rm SM}$.
The Lagrangian responsible for the lepton masses is then
\begin{eqnarray} \label{LmIII}
{\cal L}_{\rm m}^{\rm III} \,\,=\,\,
- (Y_e)_{kl}^{}\,\overline{L}_{k,L\,}^{}E_{l,R\,}^{}H \,-\,
\sqrt2\;(Y_\Sigma)_{kl}^{}\,\overline{L}_{k,L\,}^{}\Sigma_{l\,}^{}\tilde H \,-\,
\mbox{$\frac{1}{2}$}\,
(M_\Sigma)_{kl}^{}\,{\rm Tr}\bigl(\overline{\Sigma_k^{\rm c}}\,\Sigma_l^{}\bigr)
\;+\; {\rm H.c.}
\end{eqnarray}
where $\Sigma_k^{\rm c}$ is the charge conjugate of $\Sigma_k^{}$.
For convenience, we define the right-handed fields
\,${\cal E}_{k,R}^{}=\Sigma_k^-$\, and \,${\cal N}_{k,R}^{}=\Sigma_k^0$\,
and left-handed fields
\,${\cal E}_{k,L}^{}=\bigl(\Sigma_k^+\bigr)\raisebox{1pt}{$^{\rm c}$}$\, and
\,${\cal N}_{k,L}^{}=\bigl(\Sigma_k^0\bigr)\raisebox{1pt}{$^{\rm c}$}$.\,
In terms of matrices containing them and SM leptons, one can express the mass terms in
${\cal L}_{\rm m}^{\rm III}$ after electroweak symmetry breaking as
\begin{eqnarray} \label{type3mass}
{\cal L}_{\rm m}^{\rm III} \,\,\supset\,\,
-\bigl(\overline{E}_L^{} ~~~ \overline{\cal E}_L^{} \bigr)
\left(\!\begin{array}{cc} M_\ell^{} & \sqrt2\,M_{\rm D}^{} \vspace{2pt} \\
0 & M_\Sigma^{} \end{array}\!\right)
\left(\!\begin{array}{c} E_R^{} \vspace{2pt} \\ {\cal E}_R^{} \end{array}\!\right)
-\, \mbox{$\frac{1}{2}$}\, \bigl(\overline{\nu}_L^{\,\prime} ~~~ \overline{\cal N}_L^{} \bigr)
\left(\!\begin{array}{cc} 0 & M_{\rm D}^{} \vspace{2pt} \\
M_{\rm D}^{\rm T} & M_\Sigma^{} \end{array}\!\right)
\left(\!\begin{array}{c} (\nu_L')\mbox{$^{\rm c}$} \vspace{2pt} \\
{\cal N}_R^{} \end{array}\!\right)
\,+\; {\rm H.c.} \,, ~~
\end{eqnarray}
where \,$M_\ell=v Y_e/\sqrt2$\, and \,$M_{\rm D}=v Y_\Sigma/\sqrt2$\, are 3$\times$3
matrices and \,$\nu_L'=U_{\scriptscriptstyle\rm PMNS\,}^{}\nu_L^{}$.\,
For \,$M_\Sigma\gg M_{\rm D}$\, in their nonzero elements, a seesaw mechanism like that in
type I becomes operational to generate the light neutrinos' mass matrix
\begin{eqnarray}
m_\nu \,\,=\,\, -\frac{v^2}{2}\,Y_\Sigma^{}M_\Sigma^{-1}Y_\Sigma^{ T} \,.
\end{eqnarray}
Hence it is tempting simply to write $Y_\Sigma$ in a similar way to $Y_\nu$ in type I,
\begin{eqnarray} \label{Ys}
Y_\Sigma^{} \,\,=\,\,
\frac{i\sqrt2}{v}\,U_{\scriptscriptstyle\rm PMNS\,}^{}\hat m^{1/2}_\nu O M_\Sigma^{1/2} \,,
\end{eqnarray}
and use $Y_e$ in Eq.\,(\ref{Ye}) like before.

One, however, needs to justify this approximation because the light charged leptons, $E_k$,
mix with the heavy ones, ${\cal E}_k$, as can be deduced from Eq.\,(\ref{type3mass}).
They are related to the mass eigenstates $E_k'$ and ${\cal E}_k'$ by
\begin{eqnarray}
\left( \begin{array}{c} E_{\sf C}^{} \vspace{2pt} \\ {\cal E}_{\sf C}^{} \end{array} \right)
= \left(\begin{array}{ccc} (U_{EE})_{\sf C}^{} & & (U_{E\cal E})_{\sf C}^{} \vspace{2pt} \\
(U_{{\cal E}E})_{\sf C}^{} & & (U_{\cal EE})_{\sf C}^{} \end{array} \right)
\left( \begin{array}{c} E_{\sf C}' \vspace{2pt} \\ {\cal E}_{\sf C}' \end{array} \right) ,
~~~~~~~ {\sf C} \,\,=\,\, L,R \;.
\end{eqnarray}
This alters $U_{\scriptscriptstyle\rm PMNS}$ in Eq.\,(\ref{Ys})
to \,$(U_{EE})_L^\dagger U_{\scriptscriptstyle\rm PMNS}^{}$\, as well as $Y_e$
to \,$(U_{EE})_L^\dagger Y_{e\,}^{} (U_{EE})_R^{}$.\,
At leading order~\cite{Abada:2008ea}, \,$(U_{EE})_L=\openone-M_D^{}M^{-2}_\Sigma M_D^\dagger$\,
and \,$(U_{EE})_R=\openone$\, for \,$M_D\ll M_\Sigma$.\,
Thus, the deviations of $(U_{EE})_{L,R}$ from the unit matrix are negligible, and
the approximation of $Y_\Sigma$ in Eq.\,(\ref{Ys}) is justified.

Since $M_\Sigma$ in Eq.\,(\ref{LmIII}) would introduce an additional source of flavor
violation if the fermion triplets had unequal masses, in our treatment of type III with MFV,
like in type\,\,I, we restrict ourselves to the less complicated possibility that these
fermions are degenerate, \,$M_\Sigma={\cal M}\openone$.\,
Accordingly, here we only need to work with
\begin{eqnarray} \label{AIII}
{\sf A} \,\,=\,\, Y_\Sigma^{}Y^\dagger_\Sigma \,\,=\,\,
\frac{2\cal M}{v^2}\, U_{\scriptscriptstyle\rm PMNS\,}^{} \hat m^{1/2}_\nu O
O^\dagger\hat m^{1/2}_\nu U_{\scriptscriptstyle\rm PMNS}^\dagger
\end{eqnarray}
and, as in the previous scenarios, $\sf B$ in Eq.\,(\ref{ABD}), which are no different from
those in type I, where $O$ may be real or complex\,\,\cite{Cirigliano:2005ck,Branco:2006hz}.
Also, we fix the biggest eigenvalue of $\sf A$ to one.

\section{Leptonic dipole operators in seesaw models with MFV\label{dipole}}

Having set up the basics of the MFV realizations in the minimal seesaw models of types I, II,
and III, we now study some of the phenomenological implications and point out possible
differences among them.
It is clear from the last section that as far as MFV phenomenology is concerned type I and
type III will be virtually alike because, with the new fermion masses being far above the TeV
level, the building blocks for the quantity $\Delta$ are the same in both cases.
However, within the MFV context we expect that marked differences can materialize between
these two models and type II.

To explore the phenomenological consequences of MFV, one can adopt an effective theory
approach~\cite{D'Ambrosio:2002ex,Cirigliano:2005ck}, assuming that the heavy degrees of
freedom in the full theory have been integrated out.
This is especially suitable for the seesaw scenarios under consideration, where the masses
of the new particles are much greater than the energies of the processes which we examine
in this paper.
A~number of higher-dimensional effective operators involving leptons have been listed
in the leptonic MFV literature~\cite{Cirigliano:2005ck,Cirigliano:2006su}.
Here we focus on dimension-six operators which generate dipole
interactions between the SM leptons and electroweak gauge bosons.
We deal with several other leptonic operators in the next section.

The dipole operators of interest are~\cite{Cirigliano:2005ck}
\begin{eqnarray} \label{operators}
O^{(e1)}_{RL} \,=\, g'\overline{E}_R^{}Y^\dagger_e\Delta_{\ell1}^{}\sigma_{\kappa\omega}^{}
H^\dagger L_L^{}B^{\kappa\omega} ~, & ~~~~~~~ &
O^{(e2)}_{RL} \,=\, g_{\,}\overline{E}_R^{}Y^\dagger_e\Delta_{\ell2\,}^{}\sigma_{\kappa\omega}^{}
H^\dagger\tau_a^{}L_L^{}W_a^{\kappa\omega} ~,
\end{eqnarray}
where $W$ and $B$ stand for the usual SU(2)$_L$$\times$U(1)$_Y$ gauge fields with
coupling constants $g$ and~$g'$, respectively, $\tau_a^{}$ are Pauli matrices,
summation over \,$a=1,2,3$\, is implicit, and $\Delta_{\ell 1,\ell 2}$ have
the same form as $\Delta$ in Eq.\,(\ref{general}), but with generally different $\xi_r^{}$.
One can write the effective Lagrangian for $O^{(e1,e2)}_{RL}$ as
\begin{eqnarray} \label{Leff}
{\cal L}_{\rm eff}^{} \,\,=\,\,
\frac{1}{\Lambda^2} \Bigl[O^{(e1)}_{RL} + O^{(e2)}_{RL}\Bigr] \;+\; {\rm H.c}. ~,
\end{eqnarray}
where $\Lambda$ is the scale of MFV and their own coefficients in this Lagrangian have been
absorbed by the $\xi_r^{}$ in their respective $\Delta$'s.

The terms in Eq.\,(\ref{Leff}) with the photon have the general form
\begin{eqnarray} \label{Ll2l'g}
{\cal L}_{E_k^{}E_l^{}\gamma}^{} \,\,=\,\, \frac{\sqrt{\alpha\pi}}{\Lambda^2}\,
\overline{E}_{k\,}^{} \sigma_{\kappa\omega}^{} \Bigl\{
m_{E_k^{}}^{}(\Delta_\ell)_{kl}^{} + m_{E_l^{}}^{}(\Delta_\ell)_{lk}^* -
\Bigl[ m_{E_k^{}}^{}(\Delta_\ell)_{kl}^{}-m_{E_l^{}}^{}(\Delta_\ell)_{lk}^*\Bigr]\gamma_5^{}
\Bigr\} E_{l\,}^{}F^{\kappa\omega} ~,
\end{eqnarray}
where \,$\alpha\simeq1/137$\, is the fine-structure constant, $F^{\kappa\omega}$ denotes
the electromagnetic field-strength tensor, \,$(E_1,E_2,E_3)=(e,\mu,\tau)$,\, and
hereafter \,$\Delta_\ell=\Delta_{\ell1}-\Delta_{\ell2}$.\,
These interactions contribute to the flavor-changing decays
\,$E_l^{}\to E_k^{}\gamma, E_k^{}E_j^-E_j^+$\, and nuclear \,$\mu\to e$\, conversion,
as well as to the anomalous
magnetic moments ($g-2$) and electric dipole moments (EDMs) of charged leptons.
We ignore the contributions from ${\cal L}_{\rm eff}$ to \,$\mu\to e$\, conversion and
\,$E_l^{}\to E_k^{}E_j^-E_j^+$\, that are mediated by the $Z$ boson due to
the suppression by its mass.
The flavor-changing transitions and lepton EDMs have been searched for over the years,
but with null results so far, leading to increasingly severe bounds on their rates.
For the electron and muon $g-2$, the predictions and measurements have reached high precision
which continues to improve, implying that the inferred room for new physics is small and
progressively decreasing.
Thus the experimental information on these processes can offer very stringent restrictions on
the scale $\Lambda$ in Eq.\,(\ref{Leff}).
We address this in the rest of the section.

\subsection{Flavor-changing transitions and anomalous magnetic moments}

We treat first observables that are not sensitive to $CP$-violating effects.
In this case, we can retain no more than 3 of the 17 terms of $\Delta$
in Eq.\,(\ref{general}), as the others are suppressed by comparison.
Since we pick the largest eigenvalue of the $\sf A$ matrix to be one in order to enhance
the impact of new physics, the elements of $\sf A$ are much greater than those of
the $\sf B$ matrix, whose biggest eigenvalue is \,$2m_\tau^2/v^2\simeq1.0\times10^{-4}$.\,
As a consequence, the matrix elements of the terms in Eq.\,(\ref{general}) with at least one
power of $\sf B$ are in general significantly smaller due to this suppression factor than
the terms without any $\sf B$.
Thus, in dealing with such observables, we can make the approximation
\,$\Delta=\xi^{}_1\openone+\xi^{}_{2\,}{\sf A}+\xi^{}_{4\,}{\sf A}^2$,\,
leading to \,$\Delta^\dagger=\Delta$\, after the small
${\rm Im}_{\,}\xi_{1,2,4}^{}$ have been ignored.

Among the flavor-changing decays \,$E_l^{}\to E_k^{}\gamma$,\, one can expect
the most severe constraint from \,$\mu\to e\gamma$\, which has the strictest
measured limit~\cite{pdg}.
With its amplitude derived from Eq.\,(\ref{Ll2l'g}), one determines its branching ratio to be
\begin{eqnarray} \label{Bm2eg}
{\cal B}(\mu\to e\gamma) \,\,=\,\, \frac{\alpha\tau_\mu^{}m_\mu^5}{\Lambda^4}
\bigl|(\Delta_\ell)_{21}^{}\bigr|\raisebox{1pt}{$^2$} \,,
\end{eqnarray}
where $\tau_\mu^{}$ is the muon lifetime, the electron mass has been neglected, and
\begin{eqnarray} \label{D21}
(\Delta_\ell)_{jk}^{} \,\,=\,\, \xi_{2\,}^\ell{\sf A}_{jk}^{}
\,+\, \xi_{4\,}^\ell ({\sf A}\raisebox{-0.7pt}{$^2$})_{jk}^{} \,.
\end{eqnarray}
From Eq.\,(\ref{Bm2eg}), one can easily obtain the corresponding formulas for
\,$\tau\to e\gamma$\, and \,$\tau\to\mu\gamma$\, by appropriately changing the flavor indices.
These tau decays can place complementary restraints on the dipole couplings.

Searches for \,$\mu\to e$\, conversion in nuclei can offer constraints on new physics
competitive to those from \,$\mu\to e\gamma$\, measurements~\cite{deGouvea:2013zba}.
Assuming that the flavor-changing transition is again due to the dipole operators alone,
one can write the conversion rate in nucleus ${\cal N}$ as~\cite{Kitano:2002mt}
\begin{eqnarray} \label{Bm2e}
{\cal B}(\mu {\cal N}\to e{\cal N}) \,\,=\,\, \frac{\alpha\pi_{\,}m_{\mu\,}^5
\bigl|(\Delta_\ell)_{21}^{~~}D_{\cal N}^{}\bigr|\raisebox{1pt}{$^2$}}
{\Lambda^{4\,}\omega_{\rm capt}^{\cal N}} \;,
\end{eqnarray}
where $D_{\cal N}$ is the dimensionless overlap integral for~$\cal N$ and
$\omega_{\rm capt}^{\cal N}$ the rate of $\mu$ capture in ${\cal N}$.
Based on the existing experimental limits on \,$\mu\to e$\, transition in various
nuclei~\cite{pdg} and the corresponding $D_{\cal N}$ and $\omega_{\rm capt}^{\cal N}$
values~\cite{Kitano:2002mt}, one expects that the data on \,$\mu{\rm Ti}\to e{\rm Ti}$\, and
\,$\mu{\rm Au}\to e{\rm Au}$\, may supply important restrictions.
To calculate their rates, we will employ \,$D_{\rm Ti}=0.087$,\, $D_{\rm Au}=0.189$,
\,$\omega_{\rm capt}^{\rm Ti}=2.59\times10^6/\rm s$,\,
and \,$\omega_{\rm capt}^{\rm Au}=13.07\times10^6/\rm s$\,~\cite{Kitano:2002mt}.

Another kind of flavor-changing process which receives contributions
from the interactions in Eq.\,(\ref{Ll2l'g}) is the three-body decay
\,$E_l^{}\to E_k^{}E_j^-E_j^+$.\,
If there are no other contributions, one can express its rate as
\begin{eqnarray} \label{l2lll}
\Gamma_{E_l^{}\to E_k^{}E_j^{}\bar E_j^{}}^{} \,\,=\,\,
\frac{\alpha^2 m_{E_l^{}}^{5~}\bigl|(\Delta_\ell)_{lk}^{}\bigr|\raisebox{1pt}{$^2$}}
{4\pi_{\,}\Lambda^4}\; {\cal I}_{E_l^{}\to E_k^{}E_j^{}\bar E_j^{}} \,,
\end{eqnarray}
where the $m_{E_k}$ terms in Eq.\,(\ref{Ll2l'g}) have been neglected and
the factor ${\cal I}_{E_l\to E_k E_j\bar E_j}$, from the phase-space integration,
can be calculated using formulas available in the literature~\cite{Cirigliano:2006su}.
The factors relevant to the processes we will examine are
\,${\cal I}_{\mu\to ee\bar e}=9.885$,\, ${\cal I}_{\tau\to\mu\mu\bar\mu}=3.264$,\,
${\cal I}_{\tau\to\mu e\bar e}=16.97$,\,
\,${\cal I}_{\tau\to ee\bar e}=17.41$,\, and ${\cal I}_{\tau\to e\mu\bar\mu}=3.01$.\,

For numerical analysis, we need in addition the values of the various neutrino parameters,
especially their masses and the elements of the mixing matrix $U_{\scriptscriptstyle\rm PMNS}$.
For these, we adopt the numbers quoted in Table\,\,\ref{table:nudata} from
a recent fit to global neutrino data~\cite{Capozzi:2013csa}.
Most of them depend on whether neutrino masses fall into a~normal hierarchy~(NH) or
an inverted one (IH).
Since empirical information on the absolute scale of $m_{1,2,3}^{}$ is still far from
precise~\cite{pdg}, for definiteness we set \,$m_1^{}=0$ $(m_3^{}=0)$\, in the NH (IH) case.
As for the Majorana phases $\alpha_{1,2}^{}$, there are still no data available on their values.

\begin{table}[t]
\begin{tabular}{|c||c|c|} \hline Parameter & NH & IH \\ \hline\hline
$\sin^2\theta_{12}^{}$ & $0.308\pm0.017$ & $0.308\pm0.017\vphantom{\frac{1}{2}_|^|}$
\\
$\sin^2\theta_{23}$ & $0.437^{+0.033}_{-0.023}$ &
$0.455^{+0.139}_{-0.031}\vphantom{\frac{1}{2}_|^|}$ \\
$\sin^2\theta_{13}$ & $0.0234^{+0.0020}_{-0.0019}$ &
$0.0240^{+0.0019}_{-0.0022}\vphantom{\frac{1}{2}_|^|}$ \\
$\delta/\pi$ & $1.39^{+0.38}_{-0.27}$ & $1.31^{+0.29}_{-0.33}\vphantom{\frac{1}{2}_|^|}$
\\
$\delta m^2 = m_2^2-m_1^2$ &
$\left(7.54^{+0.26}_{-0.22}\right)_{\vphantom{\int}}^{\vphantom{\int}}\times10^{-5}\;\rm eV^2$ &
$\left(7.54_{-0.22}^{+0.26}\right)\times10^{-5}\;\rm eV^2$
\\
~$\Delta m^2 = \bigl|m_3^2-\bigl(m_1^2+m_2^2\bigr)/2\bigr|\vphantom{\int_{|_|}^{|^|}}$~ &
~$(2.43\pm0.06)\times10^{-3}\;\rm eV^2$~ & ~$(2.38\pm0.06)\times10^{-3}\;\rm eV^2$~
\\ \hline
\end{tabular}
\caption{Results of a recent analysis of global three-neutrino oscillation data~\cite{Capozzi:2013csa},
in terms of best-fit values and allowed 1$\sigma$ ranges of the mass-mixing parameters.
The neutrino mass hierarchy may be normal $\bigl(m_1^{}<m_2^{}<m_3^{}\bigr)$ or inverted
$\bigl(m_3^{}<m_1^{}<m_2^{}\bigr)$.\label{table:nudata}}
\end{table}

To proceed, we also need to specify the $\sf A$ matrix, which is model dependent,
as seen in the preceding section.
For the type-I or -III seesaw scenario, $\sf A$ in Eq. (\ref{AI}) [or (\ref{AIII})]
can have many different realizations, depending on $M_\nu$ and $O$.
We consider first the least complicated possibility that the right-handed neutrinos
$\nu_{k,R}$ are degenerate, with \,$M_\nu={\cal M}\openone$,\, and $O$ is a~real
orthogonal matrix, in which case
\begin{eqnarray} \label{AI'}
{\sf A} \,\,=\,\, \frac{2\cal M}{v^2}\,U_{\scriptscriptstyle\rm PMNS\,}^{}
\hat m_{\nu\,}^{}U_{\scriptscriptstyle\rm PMNS}^\dagger
\end{eqnarray}
with eigenvalues \,$\hat{\sf A}_{1,2,3}^{}=2{\cal M}m_{1,2,3}^{}/v^2$.\,
Explicitly, including this in Eq.\,(\ref{D21}) yields
\begin{eqnarray} \label{D21I}
(\Delta_\ell)_{21} &\,=\,& -c_{12\,}^{} c_{13\,}^{}
\Bigl(s_{12\,}^{}c_{23}^{}+e^{i\delta\,}c_{12\,}^{}s_{23\,}^{}s_{13}^{}\Bigr)
\Bigl( \xi_{2\,}^\ell\hat{\sf A}_1^{}+\xi_{4\,}^\ell\hat{\sf A}_1^2 \Bigr)
\nonumber \\ && \! +\;
s_{12\,}^{}c_{13\,}^{}\Bigl(c_{12\,}^{}c_{23}^{}-e^{i\delta\,}s_{12\,}^{}s_{23\,}^{}s_{13}^{}
\Bigr) \Bigl( \xi_{2\,}^\ell\hat{\sf A}_2^{}+\xi_{4\,}^\ell\hat{\sf A}_2^2 \Bigr)
\nonumber \\ && \! +\;
e^{i\delta\,}s_{23\,}^{}c_{13\,}^{}s_{13}^{}\,
\Bigl(\xi_{2\,}^\ell\hat{\sf A}_3^{}+\xi_{4\,}^\ell\hat{\sf A}_3^2\Bigr)_{\vphantom{\int}}^{} \,,
\\ \label{D31I}
(\Delta_\ell)_{31} &\,=\,& c_{12\,}^{} c_{13\,}^{}
\Bigl(s_{12\,}^{}s_{23}^{}-e^{i\delta\,}c_{12\,}^{}c_{23\,}^{}s_{13}^{}\Bigr)
\Bigl( \xi_{2\,}^\ell\hat{\sf A}_1^{}+\xi_{4\,}^\ell\hat{\sf A}_1^2 \Bigr)
\nonumber \\ && \! -\;
s_{12\,}^{}c_{13\,}^{}\Bigl(c_{12\,}^{}s_{23}^{}+e^{i\delta\,}s_{12\,}^{}c_{23\,}^{}s_{13}^{}
\Bigr) \Bigl( \xi_{2\,}^\ell\hat{\sf A}_2^{}+\xi_{4\,}^\ell\hat{\sf A}_2^2 \Bigr)
\nonumber \\ && \! +\;
e^{i\delta\,}c_{23\,}^{}c_{13\,}^{}s_{13}^{}\,
\Bigl(\xi_{2\,}^\ell\hat{\sf A}_3^{}+\xi_{4\,}^\ell\hat{\sf A}_3^2\Bigr)_{\vphantom{\int}}^{} \,,
\\ \label{D32I}
(\Delta_\ell)_{32} &\,=\,&
-\Bigl(s_{12\,}^{}c_{23}^{}+e^{-i\delta\,}c_{12\,}^{}s_{23\,}^{}s_{13}^{}\Bigr)
\Bigl(s_{12\,}^{}s_{23}^{}-e^{i\delta\,}c_{12\,}^{}c_{23\,}^{}s_{13}^{}\Bigr)
\Bigl( \xi_{2\,}^\ell\hat{\sf A}_1^{}+\xi_{4\,}^\ell\hat{\sf A}_1^2 \Bigr)
\nonumber \\ && \! -\;
\Bigl(c_{12\,}^{}c_{23}^{}-e^{-i\delta\,}s_{12\,}^{}s_{23\,}^{}s_{13}^{}\Bigr)
\Bigl(c_{12\,}^{}s_{23}^{}+e^{i\delta\,}s_{12\,}^{}c_{23\,}^{}s_{13}^{}
\Bigr) \Bigl( \xi_{2\,}^\ell\hat{\sf A}_2^{}+\xi_{4\,}^\ell\hat{\sf A}_2^2 \Bigr)
\nonumber \\ && \! +\;
c_{23\,}^{}s_{23\,}^{}c_{13}^2\,
\Bigl( \xi_{2\,}^\ell\hat{\sf A}_3^{}+\xi_{4\,}^\ell\hat{\sf A}_3^2 \Bigr) \,.
\end{eqnarray}
Later on we will also provide examples for a case in which $O$ is complex.

With $(\Delta_\ell)_{kl}$ specified, we can determine lower limits on
the MFV scale $\Lambda$ from the experimental information on the observables described above.
The particular data we use are listed in the first two columns of Table$\;$\ref{table:lfvdata}.
Since in our model-independent approach $\xi_{2,4}^{}$ are free constants,
for simplicity we assume that only one of them is nonzero at a time.
For \,$\xi_4^{}=0$,\, we scan the ranges of the neutrino mass and mixing parameters quoted in
Table$\;$\ref{table:nudata} in order to maximize $(\Delta_\ell)_{21,32,31}$ separately,
while ensuring that the values of $\cal M$ make
\,${\rm max}\bigl(\hat{\sf A}_1,\hat{\sf A}_2,\hat{\sf A}_3\bigr)=1$.\,
This allows us extract the maximal lower-limits on
\,$\hat\Lambda=\Lambda/\bigl|\xi_2^\ell\bigr|{}^{1/2}$\, from
the measured bounds on ${\cal B}(\mu\to e\gamma)$, ${\cal B}(\tau\to\mu\gamma)$, and
${\cal B}(\tau\to e\gamma)$ listed in Table$\;$\ref{table:lfvdata}, which are the strictest
ones to date in their respective groups of processes with the same flavor changes.
Subsequently, we apply the acquired values of $(\Delta_\ell)_{21,32,31}$ to obtain
limits from the other experimental bounds in this table.
We collect the $\hat\Lambda$ numbers in the third column which correspond to the NH (IH) of
neutrino masses.
For comparison, employing the central values in Table$\;$\ref{table:nudata} would give us
results which are smaller by up to 30\%.
It is worth mentioning that in all this computation the right-handed neutrino mass is
\,${\cal M}\simeq\mbox{(6.0-6.3)}\times10^{14}\;$GeV.\,

\begin{table}[b] \vspace{1ex}
\begin{tabular}{|c|c||c|c|} \hline
\multirow{2}{*}{Observable} &
\multirow{2}{*}{\raisebox{-2ex}
{$\stackrel{\displaystyle\rm Experimental}{\rm upper~bound\vphantom{|^|}}$}} &
\multicolumn{2}{c|}{$\vphantom{|_|^{\int_|}}\hat\Lambda_{\rm min}/$TeV} \\ \cline{3-4}
& &\footnotesize ~Types I and III~ &\footnotesize Type II$\vphantom{|_|^|}$ \\ \hline\hline
${\cal B}(\mu\to e\gamma)\vphantom{\frac{1}{2}_|^|}$ & $5.7\times10^{-13}$ \cite{pdg} &
338~(307) & \, 294~(312) \,
\\
${\cal B}(\mu{\rm Ti}\to e{\rm Ti})\vphantom{\frac{1}{2}_|^|}$ &
~$6.1\times10^{-13}$ \cite{Papoulias:2013gha}~ & 85~(77) & 73~(78)
\\
${\cal B}(\mu{\rm Au}\to e{\rm Au})\vphantom{\frac{1}{2}_|^|}$ &
$7.0\times10^{-13}$ \cite{pdg} & 80~(73) & ~70~(74)~
\\
${\cal B}(\mu^-\to e^-e^-e^+)\vphantom{\frac{1}{2}_|^|}$ &
$1.0\times10^{-12}$ \cite{pdg} & 81~(74) &  70~(75)
\\ \hline
${\cal B}(\tau\to\mu\gamma)\vphantom{\frac{1}{2}_|^|}$ & $4.4\times10^{-8}$ \cite{pdg} &
22~(24) & 23~(23)
\\
~${\cal B}(\tau^-\to\mu^-\mu^-\mu^+)\vphantom{\frac{1}{2}_|^|}$~ &
$2.1\times10^{-8}$ \cite{pdg} & 5.6~(5.9) & 5.9~(5.9)
\\
${\cal B}(\tau^-\to\mu^-e^-e^+)\vphantom{\frac{1}{2}_|^|}$ &
$1.8\times10^{-8}$ \cite{pdg} & 8.7~(9.3) & 9.2 (9.3)
\\ \hline
${\cal B}(\tau\to e\gamma)\vphantom{\frac{1}{2}_|^|}$ & $3.3\times10^{-8}$ \cite{pdg} &
15~(13) & 13~(13)
\\
${\cal B}(\tau^-\to e^-e^-e^+)\vphantom{\frac{1}{2}_|^|}$ &
$2.7\times10^{-8}$ \cite{pdg} & 4.9~(4.2) & 4.3~(4.2)
\\
${\cal B}(\tau^-\to e^-\mu^-\mu^+)\vphantom{\frac{1}{2}_|^|}$ &
$2.7\times10^{-8}$ \cite{pdg} & 3.2~(2.7) & 2.8~(2.7)
\\ \hline
\end{tabular}
\caption{Lower limits on \,$\hat\Lambda=\Lambda/\bigl|\xi_2^\ell\bigr|{}^{1/2}$\,
associated with dipole operators $O_{RL}^{(e1,e2)}$ inferred from experimental upper-bounds
on the branching ratios of flavor-violating leptonic transitions, as explained in the text.
In this and the remaining tables, the unbracketed (bracketed) results correspond to
the normal (inverted) hierarchy of light neutrino masses with
\,$m_{1(3)}=0$.\label{table:lfvdata}}
\end{table}

For the type-II scheme, $\sf A$ is given only in Eq.\,(\ref{AII}), which has eigenvalues
\,$\hat{\cal A}_{1,2,3}^{}=2m_{1,2,3}^2/v_T^2$\, and leads to
\begin{eqnarray} \label{D21II}
(\Delta_\ell)_{21} &\,=\,& -c_{12\,}^{} c_{13\,}^{}
\Bigl(s_{12\,}^{}c_{23}^{}+e^{i\delta\,}c_{12\,}^{}s_{23\,}^{}s_{13}^{}\Bigr)
\Bigl( \xi_{2\,}^\ell\hat{\cal A}_1^{}+\xi_{4\,}^\ell\hat{\cal A}_1^2 \Bigr)
\nonumber \\ && \! +\;
s_{12\,}^{}c_{13\,}^{}\Bigl(c_{12\,}^{}c_{23}^{}-e^{i\delta\,}s_{12\,}^{}s_{23\,}^{}s_{13}^{}
\Bigr) \Bigl( \xi_{2\,}^\ell\hat{\cal A}_2^{}+\xi_{4\,}^\ell\hat{\cal A}_2^2 \Bigr)
\nonumber \\ && \! +\;
e^{i\delta\,}s_{23\,}^{}c_{13\,}^{}s_{13}^{}\,
\Bigl( \xi_{2\,}^\ell\hat{\cal A}_3^{}+\xi_{4\,}^\ell\hat{\cal A}_3^2 \Bigr)
\end{eqnarray}
and its $(\Delta_\ell)_{31,32}$ counterparts, which are analogous to those
in Eqs. (\ref{D31I}) and (\ref{D32I}), respectively.
Utilising these matrix elements, we take steps similar to those elaborated in the previous
paragraph, while adjusting $v_T^{}$ such that
\,${\rm max}\bigl(\hat{\cal A}_1,\hat{\cal A}_2,\hat{\cal A}_3\bigr)=1$,\,
in order to extract from data the lower limits on $\hat\Lambda$ for~\,$\xi_4^\ell=0$.\,
We collect the results in the fourth column of~Table$\;$\ref{table:lfvdata},
which correspond to~\,$v_T^{}\sim0.07$\,eV.\,
With \,$\xi_2^\ell=0$\, instead, we obtain comparable numbers for
\,$\Lambda_{\rm min}/|\xi_4^\ell|^{1/2}$,\, specifically
\,285$\;$(320)$\;$TeV\, from the ${\cal B}(\mu\to e\gamma)$ data.
Now, since \,$2\hat{\cal A}_k^{}{\cal M}^2v_T^2=\hat{\sf A}_k^2v^4$,\, one realizes that
the numbers in the fourth column of Table$\;$\ref{table:lfvdata} are also the limits on
\,$\Lambda/|\xi_4^\ell|^{1/2}$\, in the type-I case of the last paragraph with \,$\xi_2^\ell=0$.\,

It is clear from Table$\;$\ref{table:lfvdata} that to date the most stringent
constraint on the dipole operators in Eq.\,(\ref{Leff}) comes from the measured bound on
\,$\mu\to e\gamma$\, among processes that change lepton flavor.
It is instructive to entertain the consequence of this for the calculated branching ratios
of the other transitions if other operators are absent or have only minor impact.
Thus, employing the $\hat\Lambda_{\rm min}$ numbers belonging to ${\cal B}(\mu\to e\gamma)$
in Table$\;$\ref{table:lfvdata} and the corresponding neutrino parameter values,
we compute the results listed in Table$\;$\ref{table:predictions}.
The ratios of any two of them and the relative size of any one of them with respect to
${\cal B}(\mu\to e\gamma)$ are, therefore, predictions of the particular scenario considered.
They can be checked experimentally if two or more of these processes are detected
in the future, as the presence of other operators with nonnegligible effects would likely
lead to a different set of predictions.
Since the numbers in Table$\;$\ref{table:predictions} are at least two orders of magnitude
below their current experimental bounds, it is of interest to make comparison with
the projected sensitivities of future experiments on lepton flavor violation.

\begin{table}[b] \vspace{1ex}
\begin{tabular}{|c||c|c|} \hline
\multirow{2}{*}{Observable} &
\multicolumn{2}{c|}{$\vphantom{|_|^{\int_|}}$Prediction} \\ \cline{2-3}
&\footnotesize ~Types I and III~ &\footnotesize Type II$\vphantom{|_|^|}$ \\ \hline\hline
${\cal B}(\mu{\rm Ti}\to e{\rm Ti})\vphantom{\frac{1}{2}_|^|}$ & $2.4\;(2.4)\times10^{-15}$ &
$2.4\;(2.4)\times10^{-15}$ \\
${\cal B}(\mu{\rm Au}\to e{\rm Au})\vphantom{\frac{1}{2}_|^|}$ & $2.2\;(2.2)\times10^{-15}$ &
$2.2\;(2.2)\times10^{-15}$ \\
${\cal B}(\mu^-\to e^-e^-e^+)\vphantom{\frac{1}{2}_|^|}$ & \, $3.3\;(3.3)\times10^{-15}$ \, &
\, $3.3\;(3.3)\times10^{-15}$ \, \\
${\cal B}(\tau\to\mu\gamma)\vphantom{\frac{1}{2}_|^|}$ & $7.9\;(14)\times10^{-13}$ &
$1.7\;(1.4)\times10^{-12}$ \\ \,
${\cal B}(\tau^-\to\mu^-\mu^-\mu^+)\vphantom{\frac{1}{2}_|^|}$ \, & $1.5\;(2.7)\times10^{-15}$ &
$3.3\;(2.6)\times10^{-15}$ \\
${\cal B}(\tau^-\to\mu^-e^-e^+)\vphantom{\frac{1}{2}_|^|}$ & $7.8\;(14)\times10^{-15}$ &
$1.7\;(1.3)\times10^{-14}$ \\
${\cal B}(\tau\to e\gamma)\vphantom{\frac{1}{2}_|^|}$ & $2.5\;(5.7)\times10^{-14}$ &
$8.6\;(4.6)\times10^{-14}$ \\
${\cal B}(\tau^-\to e^-e^-e^+)\vphantom{\frac{1}{2}_|^|}$ & $2.5\;(5.7)\times10^{-16}$ &
$8.7\;(4.7)\times10^{-16}$ \\
${\cal B}(\tau^-\to e^-\mu^-\mu^+)\vphantom{\frac{1}{2}_|^|}$ & $4.3\;(9.9)\times10^{-17}$ &
$15\;(8.0)\times10^{-17}$ \\ \hline
\end{tabular}
\caption{Predictions calculated from the contributions of the dipole operators alone, with
the $\hat\Lambda_{\rm min}$ numbers from the experimental bound on ${\cal B}(\mu\to e\gamma)$
in Table$\;$\ref{table:lfvdata} and the neutrino parameter values used to determine
them.\label{table:predictions}}
\end{table}

There are planned searches for \,$\mu\to e\gamma$\, with projected sensitivity
reaching a few times $10^{-14}$ within the next five years~\cite{Cavoto:2014qoa}.
If they come up empty, the predictions in Table$\;$\ref{table:predictions} will
decrease somewhat, probably by up to an order of magnitude.
Nevertheless, the prediction for \,$\mu\to3e$\, in Table$\;$\ref{table:predictions} will
likely still be testable with new experiments looking for it which will start running
in a couple of years and may be able to probe a branching ratio as low as $10^{-16}$ after
several years~\cite{CeiA:2014wea,Mori:2014aqa}.
Complementary checks may be available from upcoming searches for flavor-violating tau
decays which can improve their current empirical limits by two orders of
magnitude~\cite{CeiA:2014wea}.
Potentially severe restrictions will be supplied by future measurements on
\,$\mu\to e$\, conversion in nuclei which will begin in a~few years and are expected
to achieve sensitivity at the level of $10^{-17}$ or better
eventually~\cite{CeiA:2014wea,Mori:2014aqa}.

\begin{figure}[b]
\includegraphics[width=95mm]{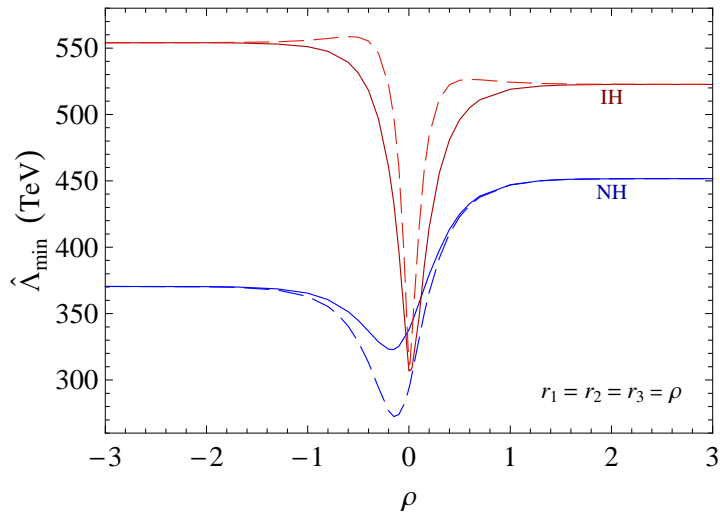}\vspace{-7pt}
\caption{Variation of the lower limit on \,$\hat\Lambda=\Lambda/|\xi_2^\ell|^{1/2}$,\, subject to
${\cal B}(\mu\to e\gamma)$ data, versus complex-$O$ parameter \,$\rho=r_1^{}=r_2^{}=r_3^{}$\, in
the absence of the Majorana phases, \,$\alpha_{1,2}^{}=0$,\, for \,$\xi_4^\ell=0$\, and
degenerate $\nu_{k,R}^{}$ (solid curves), as explained in the text.
The dashed curves depict the corresponding variation of the lower limit on
\,$\Lambda/|\xi_4^\ell|^{1/2}$\, for \,$\xi_2^\ell=0$.\label{plotrho}}
\end{figure}

As another significant observation from Table$\;$\ref{table:lfvdata}, it indicates that
no remarkable differences in the bounds on $\Lambda$ appear among the three types of seesaw
models if in type I or III the right-handed neutrinos are degenerate and the $O$ matrix is
real, with $\sf A$ in Eq.\,(\ref{AI'}).
If $O$ is complex and/or the right-handed neutrinos are nondegenerate, $\sf A$ is less
simple which may give rise to more pronounced deviations from the type-II results.
To illustrate this, we next explore the possibility that $O$ is complex.

\begin{figure}[b]
\includegraphics[width=95mm]{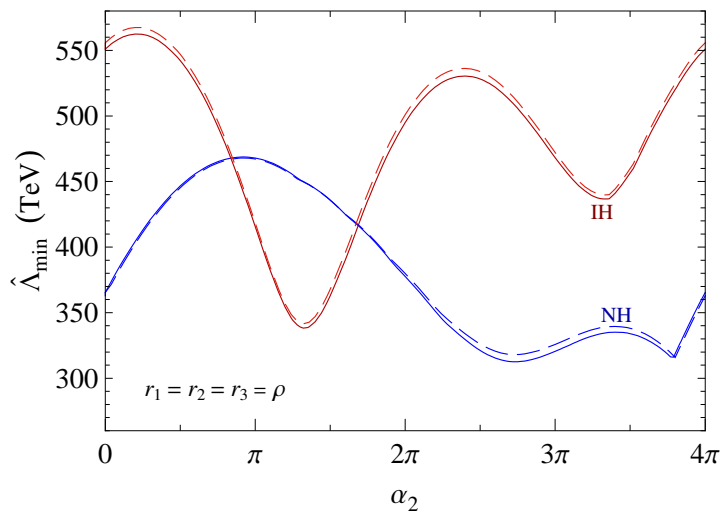}\vspace{-7pt}
\caption{Variation of the lower limit on \,$\hat\Lambda=\Lambda/|\xi_2^\ell|^{1/2}$,\, subject
to ${\cal B}(\mu\to e\gamma)$ data, versus $\alpha_2^{}$ for \,$\alpha_1^{}=0$,\,
degenerate $\nu_{k,R}^{}$, and complex-$O$ parameter \,$\rho=r_1^{}=r_2^{}=r_3^{}=-1$\,
(solid curves), as explained in the text.
The dashed curves depict the corresponding variation of the lower limit on
\,$\Lambda/|\xi_4^\ell|^{1/2}$\, for \,$\xi_2^\ell=0$.\label{plotalpha2}}
\end{figure}

With $\nu_{k,R}^{}$ still degenerate, \,$M_\nu={\cal M}\openone$,\, but $O$ complex, $\sf A$
is more complicated,
\begin{eqnarray} \label{AcomplexO}
{\sf A} \,\,=\,\, \frac{2}{v^2}\,{\cal M}_{\,}U_{\scriptscriptstyle\rm PMNS\,}^{}
\hat m^{1/2}_\nu OO^\dagger\hat m^{1/2}_\nu U_{\scriptscriptstyle\rm PMNS}^\dagger \;.
\end{eqnarray}
One can always write \,$OO^\dagger=e^{2i\sf R}$\, with a~real antisymmetric matrix
\begin{eqnarray}
{\sf R} \,\,=\, \left ( \begin{array}{ccc} 0 & r_1^{} & r_2^{} \vspace{1pt} \\
-r_1^{} & 0 & r_3^{} \vspace{1pt} \\ -r_2^{} & ~ \mbox{$-r_3^{}$} ~ & 0 \end{array} \right) .
\end{eqnarray}
Since $OO^\dagger$ is not diagonal, $\sf A$ generally has dependence on the Majorana phases
in $U_{\scriptscriptstyle\rm PMNS}$ if they are not zero.
To concentrate first on demonstrating how $O$ can bring about significant new effects,
we switch off the Majorana phases,~\,$\alpha_{1,2}^{}=0$.\,
Subsequently, for illustration, we pick \,$r_1^{}=r_2^{}=r_3^{}=\rho$\, and again
scan the parameter ranges in Table$\;$\ref{table:nudata} in order to get the highest
\,$\Lambda/|\xi_{2,4}^\ell|^{1/2}$\, from the experimental bound on ${\cal B}(\mu\to e\gamma)$,
with the condition that the largest eigenvalue of $\sf A$ in Eq.\,(\ref{AcomplexO}) is unity.
We exhibit the resulting dependence on $\rho$ in Figure$\;$\ref{plotrho}.
It reveals that, in this example, the contribution of $O$ can boost $\Lambda_{\rm min}$ by
up to 80\% with respect to its value when $O$ is real, which implies that the predicted
branching ratio is enhanced by an order of magnitude.

Turning our attention now to the impact of the Majorana phases, we make the same
choice of \,$r_{1,2,3}^{}=\rho$\, as in the preceding paragraph, select \,$\alpha_1^{}=0$\,
and \,$\rho=-1$,\, and let $\hat\Lambda_{\rm min}$ vary as a~function of $\alpha_2^{}$,
in similar steps to those in the last example.
We display the variation in~Figure$\;$\ref{plotalpha2}, which shows that although the Majorana
phases in this instance can increase the lower limits on $\Lambda$ only moderately, they can
induce sizable reduction of it.
Hence the associated decay rate is affected in roughly the same way.
All these examples demonstrate that the $O$ matrix and Majorana phases in types I and III can
produce striking effects which do not occur in type II.

Before finishing this subsection, we examine limitations from the anomalous magnetic moments,
$a_\ell^{}$, of charged leptons.
The Lagrangian for $a_\ell^{}$ is
\,${\cal L}_{a_\ell}=\sqrt{\alpha\pi}\,a_\ell^{}\,
\overline{\ell}\sigma^{\kappa\omega}\ell F_{\kappa\omega}/\bigl(2m_\ell^{}\bigr)$,\,
which gets contributions from the flavor-diagonal couplings in Eq.\,(\ref{Ll2l'g}).
Accordingly
\begin{eqnarray} \label{al}
a_{E_k}^{} \,\,=\,\, \frac{4_{\,}m_{E_k}^2}{\Lambda^2}\, \bigl[ \xi_{1\,}^\ell\delta_{kk} +
\xi_{2\,}^\ell{\sf A}_{kk}^{}+\xi_{4\,}^\ell({\sf A}\raisebox{-0.7pt}{$^2$})_{kk}^{}\bigr] \,,
\end{eqnarray}
where as earlier we have ignored the tiny ${\rm Im}_{\,}\xi_{1,2,4}^\ell$.
Since $a_e^{}$ is much suppressed by the electron mass, and since the measurement of
$a_\tau^{}$ is not yet precise~\cite{pdg}, the strongest restrictions from
anomalous magnetic moments can be expected from $a_\mu^{}$.
Currently its experimental and SM values differ by
\,$a_\mu^{\rm exp}-a_\mu^{\scriptscriptstyle\rm SM}=(249\pm87)\times10^{-11}$ \cite{Aoyama:2012wk},
which suggests that we can require the new contributions to satisfy
\,$\bigl|a_\mu^{}\bigr|<3.4\times10^{-9}$.\,
Assuming again that the right-handed neutrinos are degenerate and the $O$ matrix is real, if
only one of $\xi_{1,2,4}^\ell$ is nonzero at a time, from this $a_\mu$ bound we infer
\,$\Lambda/|\xi_{1,2,4}^\ell|^{1/2}>3.6,2.7,2.5$~TeV.\,
Upon comparing these numbers with those in Table$\;$\ref{table:lfvdata}, we conclude that
the muon $g\!-\!2$ cannot at present compete with the flavor-violating leptonic transitions in
restraining $\Lambda$.

\subsection{Lepton EDMs}

The interactions in Eq.\,(\ref{Ll2l'g}) also contribute to a charged lepton's electric dipole
moment, denoted by $d_\ell$, which is a sensitive probe for the existence of
new sources of $CP$ violation, as the SM prediction is very small~\cite{He:1989xj}.
The Lagrangian for $d_\ell^{}$ is
\,${\cal L}_{d_\ell}=
-(i d_\ell/2)\overline{\ell}\sigma^{\kappa\omega}\gamma_5\ell F_{\kappa\omega}$,\,
and so
\begin{eqnarray} \label{dl}
d_{E_k} \,\,=\,\, \frac{\sqrt2\,e_{\,}v}{\Lambda^2}\;
{\rm Im}\bigl(Y^\dagger_e \Delta_{\ell}^{}\bigr)_{kk} \;.
\end{eqnarray}
Unlike the observables treated in the previous subsection which are dominated by
a few of the terms in $\Delta_{\ell}$ with lowest orders in the $\sf A$ and $\sf B$ matrices,
the contributions pertinent to lepton EDMs are those from higher orders.
Thus, for the electron we have \cite{He:2014fva}
\begin{eqnarray} \label{de}
d_e \,\,=\,\, \frac{\sqrt2\,e_{\,}v}{\Lambda^2} \Bigl[ \xi^\ell_{12}\,
{\rm Im}\bigl(Y^\dagger_e {\sf ABA}{}^2\bigr)_{11} +
\xi^\ell_{16}\,{\rm Im}\bigl(Y^\dagger_e {\sf AB}{}^2{\sf A}{}^2\bigr)_{11} \Bigr] \,,
\end{eqnarray}
where we have again neglected ${\rm Im}_{\,}\xi^\ell_r$.
Hereafter we drop the $\xi^\ell_{16}$ part which is suppressed by one factor of $\sf B$
relative to the $\xi^\ell_{12}$ term.
The very small number of them serves to illustrate the benefit of
the resummation of the infinite series in $\Delta_{\ell}$ into the 17 terms,
as in\,\,Eq.\,(\ref{general}).

The latest analysis on $d_e^{}$ under the MFV framework has been performed in
Ref.\,\cite{He:2014fva} for the Dirac neutrino case as well as the type-I (and, hence, also
type-III) seesaw model.
If neutrinos are Dirac particles, $d_e$ has the form
\begin{eqnarray} \label{dedirac}
d_e^{\rm D} \,\,=\,\, \frac{32 e_{\,}m_e^{}}{\Lambda^2 v^8} \bigl(m_\mu^2-m_\tau^2\bigr)
\bigl(m^2_1-m_2^2\bigr) \bigl(m_2^2-m_3^2\bigr) \bigl(m_3^2-m_1^2\bigr)\,
\xi_{12\,}^\ell J_\ell^{} \;,
\end{eqnarray}
where \,$J_\ell={\rm Im}\bigl(U_{e2}^{}U_{\mu 3}^{}U_{e3}^*U^*_{\mu2}\bigr)$\, is a Jarlskog
invariant for $U_{\scriptscriptstyle\rm PMNS}$.
This turns out to be independent of the $m_{1,2,3}^{}$ individually because the neutrino
squared-mass differences defined in Table\,\,\ref{table:nudata} imply that
\,$\bigl(m^2_1-m_2^2\bigr)\bigl(m_2^2-m_3^2\bigr)\bigl(m_3^2-m_1^2\bigr)=
\delta m^2\bigl(\Delta m^2\bigr)\raisebox{1pt}{$^2$}
- \frac{1}{4}\bigl(\delta m^2\bigr)\raisebox{1pt}{$^3$}$.\,
On the other hand, in types I and III with degenerate $\nu_{k,R}^{}$ and
a real $O$ matrix, in which case $\sf A$ is given by Eq.\,(\ref{AI'}),
\begin{eqnarray} \label{dem}
d_e^{\rm I,III} \,\,=\,\, \frac{32e_{\,}m_{e\,}^{}{\cal M}^3}{\Lambda^2 v^8}
\bigl(m_\mu^2-m_\tau^2\bigr) \bigl(m_1^{}-m_2^{}\bigr)\bigl(m_2^{}-m_3^{}\bigr)
\bigl(m_3^{}-m_1^{}\bigr)\,\xi_{12\,}^\ell J_\ell^{} \;,
\end{eqnarray}
Since \,$m_k^{}\ll\cal M$,\, one can see that $d_e^{\rm D}$ is considerably suppressed
relative to $d_e^{\rm M}$.
In contrast, for type II one derives
\begin{eqnarray} \label{eq:IIedm}
d_e^{\rm II} \,\,=\,\, \frac{32e_{\,}m_e^{}}{\Lambda^2 v^2 v_T^6}\bigl(m_\mu^2-m_\tau^2\bigr)
\bigl(m_1^2-m_2^2\bigr)\bigl(m_2^2-m_3^2\bigr)\bigl(m_3^2-m_1^2\bigr)\,\xi_{12\,}^\ell J_\ell^{}
\,\,=\,\, \biggl(\frac{v}{v_T^{}}\biggr)^{\!\!6} d_e^{\rm D} \,,
\end{eqnarray}
which is far above $d_e^{\rm D}$ due to \,$v_T^{}\ll v$.\,
From these formulas, one can readily find those for $d_{\mu,\tau}$ by cyclically changing
the mass subscripts.

Numerically, \,$d_e^{\rm D}=1.3\times10^{-99\;}e\,{\rm cm\,(GeV}/\hat\Lambda)^2$\,
\cite{He:2014fva}, which is too minuscule to yield any useful restraint on
$\hat\Lambda$ from the newest data
\,$|d_e|_{\rm exp}<8.7\times10^{-29\;}e$\,cm\, from the ACME experiment~\cite{acme}.
In the Majorana neutrino case, the type-I (or -III) prediction in Eq.\,(\ref{dem})
has been evaluated in Ref.\,\cite{He:2014fva} to yield the limit
\,$\hat{\Lambda}>0.36\,(0.12)\;{\rm TeV}$\, corresponding to
\,${\cal M}=6.16\,(6.22)\times 10^{14}$\,GeV\, for the NH (IH) of neutrino masses.

In type II, from Eq.\,(\ref{eq:IIedm}) we arrive at
\begin{eqnarray}
\frac{d_e^{\rm II}}{e\,\rm cm} \,\,=\,\, 2.7\;(2.6) \times10^{-31}
\biggl(\frac{\rm eV}{v_T^{}}\biggr)^{\!\!6} \biggl(\frac{\rm GeV}{\hat\Lambda}\biggr)^{\!\!2}
\end{eqnarray}
for the NH (IH) case.
Then \,$|d_e|_{\rm exp}<8.7\times10^{-29\;}e$\,cm\, translates into
\begin{eqnarray}
\hat\Lambda \,\,>\,\, 0.055\;(0.054)\;{\rm GeV}\biggl(\frac{\rm eV}{v_T^{}}\biggr)^{\!\!3} .
\end{eqnarray}
With \,$v_T^{}\simeq0.069$\,eV\, from the requirement that the largest eigenvalue of ${\sf A}$
in Eq.\,(\ref{AII}) be unity, it follows that
\begin{eqnarray}
\hat\Lambda \,\,>\,\, 0.17\;\rm TeV \,,
\end{eqnarray}
which is roughly comparable to its counterparts in type I (or III) quoted above.

In the preceding discussion, $d_e$ is caused by the $CP$-violating Dirac phase $\delta$ in
$U_{\scriptscriptstyle\rm PMNS}$, and the Majorana phases $\alpha_{1,2}^{}$ therein do not
take part.
However, if $O$ is complex, the phases in it may give rise to an extra contribution to
$d_e$ and the Majorana phases can modify it further.
As investigated in detail in Ref.\,\cite{He:2014fva}, these new $CP$-violating contributions
to $d_e^{}$ can be more important than those of $\delta$.
Such effects do not occur in type II, as $d_e^{\rm II}$ does not have dependence on $O$ or
$\alpha_{1,2}^{}$.

\section{Additional leptonic operators\label{sec:ex-op}}

Besides the dipole operators, under the MFV framework there are other dimension-six
operators, listed in Ref.\,\cite{Cirigliano:2006su}, that can arise in the three
simplest seesaw scenarios.
Focusing on operators that are purely leptonic or couple leptons to the SM
gauge and Higgs bosons, one can write
\begin{eqnarray} \label{eq:eff-lag}
{\cal L}_{\rm eff}' \,\,=\,\, \frac{1}{\Lambda^2} \sum_{m=1}^3 O_{4L}^{(m)} +
\frac{1}{\Lambda^2}\Bigl[ O_{RL}^{(e3)} \,+\, {\rm H.c.} \Bigr] +
\frac{1}{\Lambda^2} \sum_{n=1}^2 O_{LL}^{(n)} \,,
\end{eqnarray}
where
\begin{eqnarray} \label{O'}
O^{(1)}_{4L} &\,=\,&
\overline{L}_L^{}\gamma^\mu\Delta_{\scriptscriptstyle4L}^{\scriptscriptstyle(1)}L_L^{}\,
\overline{L}_L^{}\gamma_\mu^{\;\;}\Delta_{\scriptscriptstyle4L'}^{\scriptscriptstyle(1)}L_L^{} \,,
\hspace{5em} O^{(2)}_{4L} \,\,=\,\, \overline{L}_L^{}\gamma^\mu
\Delta_{\scriptscriptstyle4L\,}^{(\scriptscriptstyle2)}\tau_a^{}L_L^{}\, \overline{L}_L^{}
\gamma_\mu^{\;\;}\Delta_{\scriptscriptstyle4L'\,}^{\scriptscriptstyle(2)}\tau_a^{}L_L^{} \,, ~~~~
\nonumber \\
O^{(3)}_{4L} &\,=\,&
\overline{L}_L^{}\gamma^\mu\Delta_{\scriptscriptstyle4L}^{\scriptscriptstyle(3)}L_L^{}\,
\overline{E}_R^{~~}\gamma_\mu^{}E_R^{} \,, \hspace{6em}\,
O_{RL}^{(e3)} \,\,=\,\, ({\cal D}_\mu H)^{\dagger\,}\overline{E}_R^{}Y_e^\dagger
\Delta_{\scriptscriptstyle RL\,}^{}{\cal D}^\mu L_L^{} \,,
\nonumber \\
O^{(1)}_{LL} &\,=\,& \mbox{$\frac{i}{2}$} \bigl[ H^\dagger ({\cal D}_\mu H)
- ({\cal D}_\mu H)^\dagger H \bigr]
\overline{L}_L^{}\gamma^\mu\Delta_{\scriptscriptstyle LL}^{\scriptscriptstyle(1)}L_L^{} \,,
\nonumber \\
O^{(2)}_{LL} &\,=\,& \mbox{$\frac{i}{2}$} \bigl[ H^\dagger\tau_a^{\;\,}
({\cal D}_\mu H)-({\cal D}_\mu H)^\dagger\tau_a^{\,\;}H \bigr] \overline{L}_L^{}\gamma^\mu
\tau_a^{\;\;}\Delta_{\scriptscriptstyle LL}^{\scriptscriptstyle(2)}L_L^{} \,,
\end{eqnarray}
with ${\cal D}_\mu$ being the usual covariant derivative involving the electroweak gauge bosons.
One could also include
\,$\overline{E}_R^{}Y_e^\dagger\Delta L_L^{}\,\overline{E}_R^{}Y_e^\dagger\Delta'L_L^{}$,\,
but this is suppressed by $Y_e^2$ relative to $O^{(1,2,3)}_{4L}$.
Since in this section we are not concerned with observables that are sensitive to $CP$
violation, the $\Delta$'s in Eq.\,(\ref{O'}) each have the approximate form
\,$\Delta=\xi^{}_1\openone+\xi^{}_{2\,}{\sf A}+\xi^{}_{4\,}{\sf A}^2=\Delta^\dagger$\, like
earlier, with generally different coefficients $\xi_{1,2,4}^{}$ of their own and
${\rm Im}_{\,}\xi_{1,2,4}^{}$ having been dropped.
We will use the above set of operators for analysis.\footnote{\baselineskip=13pt%
In Ref.\,\,\cite{Grzadkowski:2010es}, the list of independent operators includes
\,$O'=H^\dagger H_{\,}\overline{E}_R Y_e^\dagger\Delta H^\dagger L_L$\,
instead of $O_{RL}^{(e3)}$ used here.
However, $O'$ turns out to be nonindependent in the operator basis adopted in our study, being
related to $O_{RL}^{(e3)}$, with \,$\Delta_{\scriptscriptstyle RL}=\Delta$,\,
and some of the other operators we consider by
\begin{eqnarray*}
2O_{RL}^{(e3)} \,+\, {\rm H.c.} &=&
\mbox{$\frac{i}{2}$}\big[(H^\dagger{\cal D}_\mu H-({\cal D}_\mu H)^\dagger H\big] \big(
\overline{L}_L^{}\gamma^\mu Y_e^{}Y_e^\dagger\Delta L_L^{} +
2\overline{E}_R^{}\gamma^\mu Y_e^\dagger\Delta Y_e^{}E_R^{} \big)
\nonumber \\ && +\;
\mbox{$\frac{i}{2}$}\big[ H^\dagger\tau_a^{}{\cal D}_\mu H-({\cal D}_\mu H)^\dagger\tau_a^{}H\bigr]
\overline{L}_L^{}\gamma^\mu Y_e^{}Y_e^\dagger\Delta\tau_a^{}L_L^{}
\nonumber \\ && +\;
\Big[ \Big(\lambda_1^{}H^\dagger H-\mbox{$\frac{1}{2}$}m_h^2\Big)
\overline{E}_R^{}Y_e^\dagger\Delta H^\dagger L_L^{}
+ \overline{E}_R^{}Y_e^\dagger\Delta L_L^{}\,\overline{E}_R^{}Y_e^\dagger L_L^{}
\nonumber \\ && ~~~~ -\mbox{$\frac{1}{4}$}\overline{E}_R^{}\sigma_{\mu\nu}^{}H^\dagger
\bigl(g\tau_a^{}W_a^{\mu\nu}+g'B^{\mu\nu}\bigr)Y_e^\dagger\Delta L_L^{} \,+\, {\rm H.c.} \Big]
\end{eqnarray*}
plus terms involving quark fields and total derivatives, where $\lambda_1$ is the Higgs
self-coupling, \,$m_h^2=\lambda_1^{}v^2$,\, and the equations of motions for SM fields have
been applied.}

The first three of the operators in Eq.\,(\ref{O'}) contribute directly to
\,$E_k^-\to E_l^-E_l^-E_l^+$\, and \,$E_k^-\to E_l^-E_j^-E_j^+$\, with \,$l\neq j$,\,
whereas the last three contribute mainly via diagrams mediated by the $Z$ boson.
Here we assume that the dipole operators $O_{RL}^{(1,2)}$ treated in the last subsection are
absent.
To see what constraints can be derived from the experimental information on these decays,\,
for simplicity we select \,$\Delta_{4L'\,}^{(1,2)}=\openone$.\,
We then find their branching ratios to be~\cite{Cirigliano:2006su}, respectively,
\begin{eqnarray}
{\cal B}\bigl(E_k^{}\to E_l^{}E_l^{}\bar E_l^{}\bigr) &\,=\,&
\frac{\tau_{E_k}^{}m_{E_k}^5}{1536_{\,}\pi^3 \Lambda^4}
\Big[ \bigl|(A_+)_{lk}^{}\bigr|^2+2\bigl|(A_-)_{lk}^{}\bigr|^2 \Bigr] \,,
\nonumber\\
{\cal B}\bigl(E_k^{}\to E_l^{}E_j^{}\bar E_j^{}\bigr) &\,=\,&
\frac{\tau_{E_k}^{}m_{E_k}^5}{1536_{\,}\pi^3 \Lambda^4}
\Big[ \bigl|(A_+)_{lk}^{}\bigr|^2+\bigl|(A_-)_{lk}^{}\bigr|^2 \Bigr] \,,
\end{eqnarray}
where the final masses have been neglected relative to the initial one, $\tau_{E_k}$ is
the lifetime of $E_k$, and the matrices $A_\pm$ are combinations of the $\Delta$'s,
\begin{eqnarray}
A_+^{} &\,=\,& \bigl[\Delta^{\scriptscriptstyle(1)}_{\scriptscriptstyle LL}
+ \Delta^{\scriptscriptstyle(2)}_{\scriptscriptstyle LL}\bigr] \sin^{2\!}\theta_{\rm W}^{}
\,+\, \Delta^{\scriptscriptstyle(3)}_{\scriptscriptstyle4L} \;,
\nonumber \\
A_-^{} &\,=\,& \bigl[\Delta^{\scriptscriptstyle(1)}_{\scriptscriptstyle LL}
+ \Delta^{\scriptscriptstyle(2)}_{\scriptscriptstyle LL}\bigr]
\Bigl(\sin^{2\!}\theta_{\rm W}^{}-\mbox{$\frac{1}{2}$}\Bigr)
\,+\, \Delta^{\scriptscriptstyle(1)}_{\scriptscriptstyle4L}
+ \Delta^{\scriptscriptstyle(2)}_{\scriptscriptstyle4L} \;,
\end{eqnarray}
with \,$\sin^{2\!}\theta_{\rm W}^{}=0.23$.\,
We have neglected the contributions to $A_\pm$ from $O_{RL}^{(e3)}$ due to
suppression by~$m_E^2/v^2$.

To illustrate the lower limits on $\Lambda$ obtainable from the data on these decays,
already quoted in Table$\;$\ref{table:lfvdata}, for definiteness we further assume that
either only $O_{4L}^{(1,2,3)}$ with
\,$\Delta^{\scriptscriptstyle(1)}_{\scriptscriptstyle4L}=
\Delta^{\scriptscriptstyle(2)}_{\scriptscriptstyle4L}=
\Delta^{\scriptscriptstyle(3)}_{\scriptscriptstyle4L}$\,
or $O_{LL}^{(1,2)}$ with
\,$\Delta^{\scriptscriptstyle(1)}_{\scriptscriptstyle LL}=
\Delta^{\scriptscriptstyle(2)}_{\scriptscriptstyle LL}$\,
are contributing at a time, with \,$\xi_4^{}=0$\, in the $\Delta$'s,
and that in type~I (or III) the $\sf A$ matrix is given by Eq.\,(\ref{AI'}).
Using the maximal ${\sf A}_{kl}$ determined earlier, we infer the lower bounds on
$\hat\Lambda$ presented in Table$\;$\ref{table:4lepton}.

Obviously, for these operators the measured limit on ${\cal B}(\mu^-\to e^-e^-e^+)$
provides the strictest constraint among the flavor-violating processes.
To see the implication of this for the predictions on the tau three-body modes,
we calculate their branching ratios with the $\hat\Lambda_{\rm min}$ numbers belonging to
\,$\mu^-\to e^-e^-e^+$\, in Table$\;$\ref{table:4lepton} and the neutrino parameter
values employed to extract them.
We display the results in Table$\;$\ref{table:4LLLpredictions}, which are larger than
their counterparts in Table$\;$\ref{table:predictions} by roughly two orders of magnitude.
This considerable variation in predictions will help make it easier to identify
the underlying physics if one or more of the flavor-violating transitions we study are
observed in the future.

\begin{table}[t]
\begin{tabular}{|c||c|c|c|c|} \hline
\multirow{3}{*}{Process} &
\multicolumn{4}{c|}{$\vphantom{|_{\int}^{\int_|}}\hat\Lambda_{\rm min}/$TeV} \\ \cline{2-5} &
\multicolumn{2}{c|}{\footnotesize ~Types I and III~} &
\multicolumn{2}{c|}{\footnotesize Type II$\vphantom{|_|^|}$} \\ \cline{2-5} &
\footnotesize$O_{4L}^{(1,2,3)}$ & \footnotesize$O_{LL}^{(1,2)}\vphantom{|_{\int_o}^{\int_|^0}}$ &
\footnotesize$O_{4L}^{(1,2,3)}$ & \footnotesize$O_{LL}^{(1,2)}$ \\ \hline\hline
$\mu^-\to e^-e^-e^+\vphantom{\frac{1}{2}_|^|}$ & \, 118~(107) \, & 64~(58) & \, 102~(109) \, &
56~(59) \\ \hline
\, $\tau^-\to\mu^-\mu^-\mu^+\vphantom{\frac{1}{2}_|^|}$ \, & 11~(11) & 5.8~(6.2) & 11~(11) &
6.1~(6.2)
\\
$\tau^-\to\mu^-e^-e^+\vphantom{\frac{1}{2}_|^|}$ & 9.6~(10) & \, 5.4~(5.7) \, &
10~(10) & \, 5.7~(5.7) \,
\\ \hline
$\tau^-\to e^-e^-e^+\vphantom{\frac{1}{2}_|^|}$ & 6.2~(5.3) & 3.4~(2.9) &
5.5~(5.3) & 3.0~(2.9)
\\
$\tau^-\to e^-\mu^-\mu^+\vphantom{\frac{1}{2}_|^|}$ & 5.4~(4.6) & 3.0~(2.6) &
4.7~(4.6) & 2.7~(2.6)
\\ \hline
\end{tabular}
\caption{Lower limits on \,$\hat\Lambda=\Lambda/|\xi_2^{}|^{1/2}$\, associated with
operators $O_{4L}^{(1,2,3)}$ and $O_{LL}^{(1,2)}$ inferred from data on flavor-violating
decays of charged leptons, as explained in the text.
Only $O_{4L}^{(1,2,3)}$ or $O_{LL}^{(1,2)}$ are assumed to be present at
a time.\label{table:4lepton}}
\end{table}

\begin{table}[b]
\begin{tabular}{|c||c|c|c|c|} \hline
\multirow{3}{*}{Observable} &
\multicolumn{4}{c|}{Prediction} \\ \cline{2-5} &
\multicolumn{2}{c|}{\footnotesize ~Types I and III~} &
\multicolumn{2}{c|}{\footnotesize Type II$\vphantom{|_|^|}$} \\ \cline{2-5} &
\footnotesize$O_{4L}^{(1,2,3)}$ & \footnotesize$O_{LL}^{(1,2)}\vphantom{|_{\int_o}^{\int_|^0}}$
& \footnotesize$O_{4L}^{(1,2,3)}$ & \footnotesize$O_{LL}^{(1,2)}$ \\ \hline\hline
\, ${\cal B}(\tau^-\to\mu^-\mu^-\mu^+)\vphantom{\frac{1}{2}_|^|}$ \, &
$1.4\;(2.5)\times10^{-12}$    &    $1.4\;(2.5)\times10^{-12}$ &
$3.0\;(2.4)\times10^{-12}$    &    $3.0\;(2.4)\times10^{-12}$
\\
${\cal B}(\tau^-\to\mu^-e^-e^+)\vphantom{\frac{1}{2}_|^|}$ &
$7.7\;(14)\times10^{-13}$     &    $8.8\;(16)\times10^{-13}$ &
$1.7\;(1.3)\times10^{-12}$    &    $1.9\;(1.5)\times10^{-12}$
\\
${\cal B}(\tau^-\to e^-e^-e^+)\vphantom{\frac{1}{2}_|^|}$ & \,
$4.3\;(9.9)\times10^{-14}$ \, & \, $4.3\;(9.9)\times10^{-14}$ \, &
$15\;(8.1)\times10^{-14}$     &    $15\;(8.1)\times10^{-14}$
\\
${\cal B}(\tau^-\to e^-\mu^-\mu^+)\vphantom{\frac{1}{2}_|^|}$ &
$2.4\;(5.5)\times10^{-14}$    &     $2.7\;(6.3)\times10^{-14}$ & \,
$8.4\;(4.5)\times10^{-14}$ \, & \,  $9.5\;(5.1)\times10^{-14}$ \,
\\ \hline
\end{tabular}
\caption{Predictions calculated from the contributions of either $O_{4L}^{(1,2,3)}$ or
$O_{LL}^{(1,2)}$ alone, with the $\hat\Lambda_{\rm min}$ numbers from the experimental bound
on ${\cal B}(\mu^-\to e^-e^-e^+)$ in Table$\;$\ref{table:4lepton} and the neutrino parameter
values used to determine them.\label{table:4LLLpredictions}}
\end{table}

The first two of the operators in Eq.\,(\ref{eq:eff-lag}) also contribute to
\,$E_k^-\to E_l^-E_l^-E_j^+$\, with \,$l\neq j$.\,
Its branching ratio is
\begin{eqnarray}
{\cal B}\bigl(E_k^{}\to E_l^{}E_l^{}\bar E_j^{}\bigr) \,\,=\,\, \frac{\tau_{E_k}^{}m_{E_k}^5}
{768_{\,}\pi^3\Lambda^4}\bigl|D_{E_k\to E_l E_l\bar E_j}\bigr|^2 \,,
\end{eqnarray}
where
\begin{eqnarray}
D_{E_k\to E_l E_l\bar E_j} \,\,=\,\,
\bigl[\Delta_{\scriptscriptstyle4L}^{\scriptscriptstyle(1)}\bigr]_{lk}
\bigl[\Delta_{\scriptscriptstyle4L'}^{\scriptscriptstyle(1)}\bigr]_{lj}
+ \bigl[\Delta_{\scriptscriptstyle4L}^{\scriptscriptstyle(1)}\bigr]_{lj}
\bigl[\Delta_{\scriptscriptstyle4L'}^{\scriptscriptstyle(1)}\bigr]_{lk}
+ \bigl[\Delta_{\scriptscriptstyle4L}^{\scriptscriptstyle(2)}\bigr]_{lk}
\bigl[\Delta_{\scriptscriptstyle4L'}^{\scriptscriptstyle(2)}\bigr]_{lj}
+ \bigl[\Delta_{\scriptscriptstyle4L}^{\scriptscriptstyle(2)}\bigr]_{lj}
\bigl[\Delta_{\scriptscriptstyle4L'}^{\scriptscriptstyle(2)}\bigr]_{lk} \;.
\end{eqnarray}
For simplicity, we choose
\,$\bigl[\Delta_{\scriptscriptstyle4L}^{\scriptscriptstyle(1)}\bigr]_{lk}
\bigl[\Delta_{\scriptscriptstyle4L'}^{\scriptscriptstyle(1)}\bigr]_{lj}=
\bigl[\Delta_{\scriptscriptstyle4L}^{\scriptscriptstyle(2)}\bigr]_{lk}
\bigl[\Delta_{\scriptscriptstyle4L'}^{\scriptscriptstyle(2)}\bigr]_{lj}=
\xi_2^{\;\;} {\sf A}_{lk}^{}{\sf A}_{lj}^{}$.\,
Subsequently, we scan the parameter ranges in Table$\;$\ref{table:nudata} to maximize
\,${\sf A}_{lk\,}{\sf A}_{lj}$,\, while setting the largest eigenvalue of $\sf A$ to unity.
With the results, we extract from the experimental bounds
\,${\cal B}(\tau^-\to e^-e^-\mu^+)<1.5\times 10^{-8}$\, and
\,${\cal B}(\tau^-\to\mu^-\mu^-e^+)<1.7\times 10^{-8}$\, \cite{pdg}, respectively, the limits
\,$\hat\Lambda_{\rm min}>2.9\;(2.7)$ and \,6.0$\;$(5.8)~TeV\, in type I or III for
the normal (inverted) hierarchy of neutrino masses.
The corresponding results for type II are \,$\hat\Lambda_{\rm min}=2.7\;(2.7)$
and \,5.5$\;$(5.8)~TeV,\, respectively.
If the above choice for these operators is to satisfy the measured limit on
${\cal B}(\mu^-\to e^-e^-e^+)$ as well, we arrive at $\hat\Lambda_{\rm min}$ in
the (27-146)\,TeV range instead and the branching ratios of
\,$\tau^-\to e^-e^-\mu^+,\mu^-\mu^-e^+$\, below $10^{-10}$,
like those in Table$\;$\ref{table:4LLLpredictions}.

We note that the renormalizable couplings of the scalar triplet to leptons as described by
Eq.\,(\ref{LmII}) induce at tree level $T$-mediated diagrams that correspond to extra
operators such as~\,$(Y_T)_{km}^{}(Y_T)_{ln\,}^*\overline{L}_{k,L}^{}\gamma^\mu L_{l,L\,}^{}
\bar L_{m,L}^{}\gamma_\mu^{\;\;}L_{n,L}^{}/M_T^2$\, \cite{Cirigliano:2005ck,Gavela:2009cd},
which we do not analyze explicitly in this work.
They also contribute to the three-body charged-lepton decays, and so for
\,$(Y_T)_{kl}=\cal O$(1) the lower bounds on $M_T$ are comparable in order of magnitude to those
on $\hat\Lambda_{\rm min}$ in Table$\;$\ref{table:4lepton}, although $M_T$ in general is
not the same as $\Lambda$.
Thus our requirement in type II that the biggest eigenvalue of \,${\sf A}=Y_T^{}Y_T^\dagger$\,
be unity translates into the limitation \,$M_T>\cal O$(100$\;$TeV)\, according to the table.
With such a mass, the triplet scalars would be undetectable at the LHC.
If we relax the condition on~$\sf A$, the minimum of $M_T$ can be lowered, but at the same
time $\hat\Lambda_{\rm min}$ also becomes weakened.
Specifically, for $M_T$ at the TeV level, which may be within LHC reach,
$Y_T$ needs to be two orders of magnitude smaller.

Finally, we address the potential impact of one of the operators in
Eq.\,(\ref{eq:eff-lag}) on the leptonic decay of the recently discovered Higgs boson.
As data from the LHC will continue to accumulate with increasing precision, they may
uncover clues of new physics in the couplings of the particle.
In general operators beyond the SM involving the Higgs boson can bring about modifications
to its standard decay modes and/or cause it to undergo exotic decays~\cite{Curtin:2013fra}.
Therefore, here we investigate to what extent this may occur due to $O_{RL}^{(e3)}$.
Although $O_{LL}^{(1,2)}$ also contain $H$, they each have no tree-level contributions to
the Higgs decay into a pair of leptons.

The latest LHC measurements have begun to reveal the Higgs couplings to charged leptons.
The ATLAS and CMS Collaborations have observed \,$h\to\tau^+\tau^-$\,
and measured its signal strength to be
\,$\sigma/\sigma_{\scriptscriptstyle\rm SM}^{}=1.42^{+0.44}_{-0.38}$
and $0.91\pm0.27$,\, respectively~\cite{CMS:2014ega,atlas:h2tt}.
In contrast, the only experimental information available on \,$h\to\mu^-\mu^+$\, to date
are the bounds \,${\cal B}(h\to\mu^-\mu^+)<1.5\times10^{-3}$ and $1.6\times10^{-3}$\,
from ATLAS and CMS, respectively~\cite{atlas:h2mm,cms:h2mm}.
On the other hand, CMS~\cite{cms:h2mt} has intriguingly reported the detection of
a slight excess of flavor-violating \,$h\to\mu^\pm\tau^\mp$\, events with
a\,\,significance of~2.5$\sigma$.
If the finding is interpreted as a statistical fluctuation, it translates into the limit
\,${\cal B}(h\to\mu\tau)={\cal B}(h\to\mu^-\tau^+)+{\cal B}(h\to\mu^+\tau^-)<1.57$\%\,
at~95\% C.L.~\cite{cms:h2mt}.
In view of these data, we demand nonstandard contributions to respect
\begin{eqnarray} \label{higgsconstraints}
0.7 \,\,<\,\,
\frac{\Gamma_{h\to\tau\bar\tau}}{\Gamma_{h\to\tau\bar\tau}^{\scriptscriptstyle\rm SM}}
\,\,<\,\, 1.8 ~, ~~~~~~~
\frac{\Gamma_{h\to\mu\bar\mu}}{\Gamma_{h\to\mu\bar\mu}^{\scriptscriptstyle\rm SM}}
\,\,<\,\, 6.7 ~, ~~~~~~~
\frac{\Gamma_{h\to\mu\tau}}{\Gamma_h^{\scriptscriptstyle\rm SM}}
\,\,<\,\, 1.5\% \,,
\end{eqnarray}
where \,$\Gamma_{h\to\tau\bar\tau}^{\scriptscriptstyle\rm SM}=257$\,keV,
\,$\Gamma_{h\to\mu\bar\mu}^{\scriptscriptstyle\rm SM}=894$\,eV,\, and
\,$\Gamma_h^{\scriptscriptstyle\rm SM}=4.08\;$MeV \cite{lhctwiki} are the SM widths
for a~Higgs mass \,$m_h^{}=125.1$\,GeV,\, which reflects the average of the newest
measurements~\cite{mhx}.

One can write the amplitude for \,$h\to E_k^-E_l^+$\, as
\begin{eqnarray}
{\cal M}_{h\to E_k^{}\bar E_l^{}} \,\,=\,\, \bar u_{E_k}^{}
\bigl(y_{kl\,}^L P_L^{}+y_{kl\,}^R P_R^{}\bigr)v_{E_l}^{} \;,
\end{eqnarray}
where $u$ and $v$ are the leptons' spinors and in the SM at tree level
\,$y_{kl}^{L,R}=\delta_{kl\,}^{}m_{E_k}/v$.\,
Its decay rate is then
\begin{eqnarray}
\Gamma_{h\to E_k^{}\bar E_l^{}} \,\,=\,\, \frac{m_h^{}}{16\pi}
\Bigl(\bigl|y_{kl}^L\bigr|^2+\bigl|y_{kl}^R\bigr|^2\Bigr) \,,
\end{eqnarray}
where in the kinematic factor the lepton masses have been neglected compared to $m_h^{}$.
Including the contribution of $O_{RL}^{(e3)}$, we have
\begin{eqnarray}
y_{kl}^L \,\,=\,\, \frac{\delta_{kl}^{}\,m_{E_k^{}}^{}}{v} \,-\,
\frac{m_{E_k^{}}^{}m_h^2}{2\Lambda^2v} (\Delta_{\scriptscriptstyle RL})_{kl}^{}
\,\,=\,\, \bigl(y_{lk}^R\bigr)^* \,.
\end{eqnarray}
Upon maximizing the relevant elements of the $\sf A$ matrix, which is in Eq.\,(\ref{AI'})
for type I and Eq.\,(\ref{AII}) for type\,\,II, we calculate the results collected in
Table$\;$\ref{table:h2ll'} for \,$\xi_1^{}=\xi_4^{}=0$,\,
which fullfil the conditions in\,\,Eq.\,(\ref{higgsconstraints}).
To get the numbers in the table, we treated the flavor-diagonal
modes independently of the $\mu^\pm\tau^\mp$ channels.
If the requirements from \,$h\to\mu^-\mu^+,\tau^-\tau^+$\, data, which led to the higher
$\hat\Lambda_{\rm min}$ values in the table, are to be satisfied also by the contributions
to\,\,\,$h\to\mu^\pm\tau^\mp$,\, we find that
$\Gamma_{h\to\mu\tau}/\Gamma_h^{\scriptscriptstyle\rm SM}$ cannot be more than
about~$0.11\%$.

\begin{table}[h] \vspace{1em}
\begin{tabular}{|c||c|c|} \hline
\multirow{2}{*}{Process} &
\multicolumn{2}{c|}{$\vphantom{|_{\int}^{\int_|}}\hat\Lambda_{\rm min}/$GeV} \\ \cline{2-3} &
\footnotesize ~Types I and III~ & \footnotesize Type II$\vphantom{|_|^|}$ \\ \hline\hline
\, $h\to\mu^-\mu^+,\tau^-\tau^+\vphantom{\frac{1}{2}_|^|}$ \, &
175~(170) & \, 168~(170) \, \\ \hline
$h\to\mu^\mp\tau^\pm\vphantom{\frac{1}{2}_|^|}$ & \, 83~(88) \, & \, 88~(87) \,
\\ \hline
\end{tabular}
\caption{Lower limits on \,$\hat\Lambda=\Lambda/|\xi_2^{}|^{1/2}$\, associated with operator
$O_{RL}^{(e3)}$ inferred from measurements on dilepton Higgs decays, as explained in
the text.\label{table:h2ll'}}
\end{table}

\section{Conclusions\label{conclusion}}

The application of the MFV hypothesis to the lepton sector provides a framework for
systematically analyzing the predictions of different models in which lepton-flavor
nonconservation and $CP$ violation arise from the leptonic Yukawa couplings.
We have explored this in the simplest seesaw scenarios where neutrino mass generation is
mediated by new fermion singlets (type I) or triplets (type III) and by a~scalar triplet (type II).
Taking a model-independent effective-theory approach, we consider the phenomenological
implications by analyzing the contributions of new interactions that are organized according
to the MFV hypothesis and consist of only a limited number of terms which have been resummed
from an infinite number of them.

More specifically, we evaluate constraints on the MFV scale $\Lambda$ associated with
leptonic dipole operators from the latest experimental information on flavor-violating
$E_k^{}\to E_l^{}\gamma$\, decays, nuclear \,$\mu\to e$\, conversion, flavor-violating
three-body decays of charged leptons, muon $g-2$, and the electron's EDM.
We find that the existing data, especially the bound on ${\cal B}(\mu\to e\gamma)$, can
restrict the lower limit on $\Lambda$ to over 500$\;$TeV or more,
depending on the details of the seesaw scheme.
In types I and III, this corresponds to the new fermions responsible for the seesaw mechanism
being super heavy, with masses roughly of order \,$10^{15}$\,GeV.\,
On the other hand, it is interesting to point out that in type II, although the VEV of
the scalar triplet needs to be \,$v_T^{}\sim0.07$\,eV\, in our approach, its mass does not
have to be \,$10^{15}$\,GeV\, and can be as low as a few hundred\,\,TeV.
If $M_T$ is to be within LHC reach, in the TeV range, $v_T^{}$ has to be of
$\cal O$(10$\;$eV) instead.
Another major difference between type I (or III) and type II is that in the former
the Yukawa couplings of the new fermions contain features which can have substantially
enhancing effects, including $CP$-violating ones, and which do not exist in type II.

Beyond the dipole operators, we look at additional dimension-six operators
that involve only leptons or couple them to the SM gauge and Higgs bosons.
Since these operators contribute to the flavor-changing three-body decays as well,
the associated MFV scale must satisfy their experimental bounds.
Based on the resulting strongest constraints, we estimate predictions on some of these processes
which are markedly distinguishable from the corresponding predictions from the dipole operators.

It is interesting that one of the extra operators can also contribute to the flavor-conserving
and flavor-violating leptonic decays of the Higgs boson and is therefore subject to constraints
from future Higgs measurements at the LHC which will continue to improve in precision.
Upcoming searches for other processes that violate lepton flavor will offer complementary
tests on the various operators in the different seesaw scenarios we have studied.
The examples we have presented serve to illustrate the great importance of making such
experimental efforts.

\acknowledgments

This research was supported in part by the MOE Academic Excellence Program (Grant No. 102R891505)
and NSC of ROC and by NSFC (Grant No. 11175115) and Shanghai Science and Technology Commission
(Grant No. 11DZ2260700) of PRC.


\begin{thebibliography}{0}

\bibitem{pdg}
  K.A.~Olive {\it et al.}  [Particle Data Group Collaboration],
  Chin.\ Phys.\ C {\bf 38}, 090001 (2014).

\bibitem{mfv1}
  R.S.~Chivukula and H.~Georgi,
  Phys.\ Lett.\ B {\bf 188}, 99 (1987);
L.J.~Hall and L.~Randall,
  Phys.\ Rev.\ Lett.\  {\bf 65}, 2939 (1990);
  A.J.~Buras, P.~Gambino, M.~Gorbahn, S.~Jager, and L.~Silvestrini,
  Phys.\ Lett.\ B {\bf 500}, 161 (2001)  [hep-ph/0007085];
  A.J.~Buras,
  Acta Phys.\ Polon.\ B {\bf 34}, 5615 (2003)  [hep-ph/0310208];
  A.L.~Kagan, G.~Perez, T.~Volansky, and J.~Zupan,
  Phys.\ Rev.\ D {\bf 80}, 076002 (2009)  [arXiv:0903.1794 [hep-ph]].

\bibitem{D'Ambrosio:2002ex}
  G.~D'Ambrosio, G.F.~Giudice, G.~Isidori, and A.~Strumia,
  Nucl.\ Phys.\ B {\bf 645}, 155 (2002)  [hep-ph/0207036].

\bibitem{Cirigliano:2005ck}
  V.~Cirigliano, B.~Grinstein, G.~Isidori, and M.B.~Wise,
  Nucl.\ Phys.\ B {\bf 728}, 121 (2005)  [hep-ph/0507001].

\bibitem{Cirigliano:2006su}
  V.~Cirigliano and B.~Grinstein,
  Nucl.\ Phys.\ B {\bf 752}, 18 (2006)  [hep-ph/0601111].

\bibitem{Branco:2006hz}
G.C.~Branco, A.J.~Buras, S.~Jager, S.~Uhlig, and A.~Weiler,
  JHEP {\bf 0709}, 004 (2007)  [hep-ph/0609067].

\bibitem{Davidson:2006bd}
  S.~Davidson and F.~Palorini,
  Phys.\ Lett.\ B {\bf 642}, 72 (2006)  [hep-ph/0607329];
  A.S.~Joshipura, K.M.~Patel, and S.K.~Vempati,
  Phys.\ Lett.\ B {\bf 690}, 289 (2010)  [arXiv:0911.5618 [hep-ph]];
  R.~Alonso, G.~Isidori, L.~Merlo, L.A.~Munoz, and E.~Nardi,
  JHEP {\bf 1106}, 037 (2011)  [arXiv:1103.5461 [hep-ph]];
  D.~Aristizabal Sierra, A.~Degee, and J.F.~Kamenik,
  JHEP {\bf 1207}, 135 (2012)  [arXiv:1205.5547 [hep-ph]].

\bibitem{Gavela:2009cd}
M.B.~Gavela, T.~Hambye, D.~Hernandez, and P.~Hernandez,
  JHEP {\bf 0909}, 038 (2009)  [arXiv:0906.1461 [hep-ph]].

\bibitem{seesaw1}
  P.~Minkowski,
  Phys.\ Lett.\  B {\bf 67}, 421 (1977);
T.~Yanagida, in {\it Proceedings of the Workshop on the Unified Theory and the Baryon Number in
the Universe}, edited by O.~Sawada and A.~Sugamoto (KEK, Tsukuba, 1979), p.~95;
  Prog.\ Theor.\ Phys.\  {\bf 64}, 1103 (1980);
M.~Gell-Mann, P.~Ramond, and R.~Slansky,
in {\it Supergravity}, edited by P.~van Nieuwenhuizen and D.~Freedman
(North-Holland, Amsterdam, 1979), p.~315;
  P.~Ramond,
  arXiv:hep-ph/9809459;
S.L.~Glashow, in {\it Proceedings of the 1979 Cargese Summer Institute on Quarks and Leptons},
edited by M.~Levy {\it et al}. (Plenum Press, New York, 1980), p.~687;
  R.N.~Mohapatra and G.~Senjanovic,
  Phys.\ Rev.\ Lett.\  {\bf 44}, 912 (1980);
J.~Schechter and  J.W.F.~Valle,
  Phys.\  Rev.\ D {\bf 25}, 774 (1982).

\bibitem{seesaw12}
J.~Schechter and  J.W.F.~Valle,
  Phys.\  Rev.\ D {\bf 22}, 2227 (1980).

\bibitem{seesaw2}
M.~Magg and C.~Wetterich,
  Phys.\ Lett.\ B {\bf 94}, 61 (1980);
T.P.~Cheng and L.F.~Li,
  Phys.\ Rev.\ D {\bf 22}, 2860 (1980);
  R.N.~Mohapatra and G.~Senjanovic,
  Phys.\ Rev.\ D {\bf 23}, 165 (1981);
  G.~Lazarides, Q.~Shafi, and C.~Wetterich,
  Nucl.\ Phys.\ B {\bf 181}, 287 (1981).

\bibitem{seesaw3}
  R.~Foot, H.~Lew, X.G.~He, and G.C.~Joshi,
  Z.\ Phys.\  C {\bf 44}, 441 (1989).

\bibitem{Colangelo:2008qp}
  G.~Colangelo, E.~Nikolidakis, and C.~Smith,
  Eur.\ Phys.\ J.\ C {\bf 59}, 75 (2009)  [arXiv:0807.0801 [hep-ph]];
  L.~Mercolli and C.~Smith,
  Nucl.\ Phys.\ B {\bf 817}, 1 (2009)  [arXiv:0902.1949 [hep-ph]].

\bibitem{He:2014fva}
  X.G.~He, C.J.~Lee, S.F.~Li, and J.~Tandean,
  Phys.\ Rev.\ D {\bf 89}, 091901 (2014)  [arXiv:1401.2615 [hep-ph]];
  JHEP {\bf 1408}, 019 (2014)  [arXiv:1404.4436 [hep-ph]].

\bibitem{Casas:2001sr}
  J.A.~Casas and A.~Ibarra,
  Nucl.\ Phys.\ B {\bf 618}, 171 (2001)   [hep-ph/0103065].

\bibitem{Gelmini:1980re}
  G.B.~Gelmini and M.~Roncadelli,
  Phys.\ Lett.\ B {\bf 99}, 411 (1981);
  H.M.~Georgi, S.L.~Glashow, and S.~Nussinov,
  Nucl.\ Phys.\ B {\bf 193}, 297 (1981).

\bibitem{Abada:2008ea}
  A.~Abada, C.~Biggio, F.~Bonnet, M.B.~Gavela, and T.~Hambye,
  Phys.\ Rev.\  D {\bf 78}, 033007 (2008)  [arXiv:0803.0481 [hep-ph]].

\bibitem{deGouvea:2013zba}
  A.~de Gouvea and P.~Vogel,
  Prog.\ Part.\ Nucl.\ Phys.\  {\bf 71}, 75 (2013)  [arXiv:1303.4097 [hep-ph]].

\bibitem{Kitano:2002mt}
  R.~Kitano, M.~Koike, and Y.~Okada,
  Phys.\ Rev.\ D {\bf 66}, 096002 (2002)  [Erratum-ibid.\ D {\bf 76}, 059902 (2007)]
  [hep-ph/0203110].

\bibitem{Capozzi:2013csa}
  F.~Capozzi, G.L.~Fogli, E.~Lisi, A.~Marrone, D.~Montanino, and A.~Palazzo,
  Phys.\ Rev.\ D {\bf 89}, 093018 (2014)  [arXiv:1312.2878 [hep-ph]].

\bibitem{Papoulias:2013gha}
D.K.~Papoulias and T.S.~Kosmas,
  Phys.\ Lett.\ B {\bf 728}, 482 (2014)  [arXiv:1312.2460 [nucl-th]].

\bibitem{Cavoto:2014qoa}
  G.~Cavoto,
  arXiv:1407.8327 [hep-ex].

\bibitem{CeiA:2014wea}
  F.~Cei and D.~Nicolo,
  Adv.\ High Energy Phys.\  {\bf 2014}, 282915 (2014).

\bibitem{Mori:2014aqa}
  T.~Mori and W.~Ootani,
  Prog.\ Part.\ Nucl.\ Phys.\  {\bf 79}, 57 (2014).

\bibitem{Aoyama:2012wk}
  T.~Aoyama, M.~Hayakawa, T.~Kinoshita, and M.~Nio,
  Phys.\ Rev.\ Lett.\  {\bf 109}, 111808 (2012)  [arXiv:1205.5370 [hep-ph]].

\bibitem{He:1989xj}
  X.G.~He, B.H.J.~McKellar, and S.~Pakvasa,
  Int.\ J.\ Mod.\ Phys.\ A {\bf 04}, 5011 (1989)  [Erratum-ibid.\ A {\bf 06}, 1063 (1991)];
W.~Bernreuther and M.~Suzuki,
  Rev.\ Mod.\ Phys.\  {\bf 63}, 313 (1991)  [Erratum-ibid.\  {\bf 64}, 633 (1992)];
  J.S.M.~Ginges and V.V.~Flambaum,
  Phys.\ Rept.\  {\bf 397}, 63 (2004)  [physics/0309054];
M.~Pospelov and A.~Ritz,
 Annals Phys.\  {\bf 318}, 119 (2005)  [hep-ph/0504231];
T.~Fukuyama,
  Int.\ J.\ Mod.\ Phys.\ A {\bf 27}, 1230015 (2012)  [arXiv:1201.4252 [hep-ph]];
J.~Engel, M.J.~Ramsey-Musolf, and U.~van Kolck,
  Prog.\ Part.\ Nucl.\ Phys.\  {\bf 71}, 21 (2013)  [arXiv:1303.2371 [nucl-th]].

\bibitem{acme}
  J.~Baron {\it et al.}  [ACME Collaboration],
  Science {\bf 343}, no. 6168, 269 (2014)
  [arXiv:1310.7534 [physics.atom-ph]].

\bibitem{Grzadkowski:2010es}
  B.~Grzadkowski, M.~Iskrzynski, M.~Misiak, and J.~Rosiek,
  JHEP {\bf 1010}, 085 (2010)  [arXiv:1008.4884 [hep-ph]].

\bibitem{Curtin:2013fra}
D.~Curtin, R.~Essig, S.~Gori, P.~Jaiswal, A.~Katz, T.~Liu, Z.~Liu and D.~McKeen {\it et al.},
  Phys.\ Rev.\ D {\bf 90}, 075004 (2014)  [arXiv:1312.4992 [hep-ph]].

\bibitem{CMS:2014ega}
CMS Collaboration,
Report No. CMS-PAS-HIG-14-009, July 2014.

\bibitem{atlas:h2tt}
ATLAS collaboration,
Report No. ATLAS-CONF-2014-061, ATLAS-COM-CONF-2014-080, October 2014.

\bibitem{atlas:h2mm}
  G.~Aad {\it et al.}  [ATLAS Collaboration],
  Phys.\ Lett.\ B {\bf 738}, 68 (2014)  [arXiv:1406.7663 [hep-ex]].

\bibitem{cms:h2mm}
  V.~Khachatryan {\it et al.}  [CMS Collaboration],
  Phys.\ Lett.\ B {\bf 744}, 184 (2015)  [arXiv:1410.6679 [hep-ex]].
  
\bibitem{cms:h2mt}
  CMS Collaboration,
Report No. CMS-PAS-HIG-14-005, July 2014.

\bibitem{lhctwiki}
  S.~Heinemeyer {\it et al.}  [LHC Higgs Cross Section Working Group Collaboration],
  arXiv:1307.1347 [hep-ph].
Updated values of the SM Higgs total width available at
https://twiki.cern.ch/twiki/bin/view/LHCPhysics/CERNYellowReportPageBR3.

\bibitem{mhx}
  G.~Aad {\it et al.}  [ATLAS Collaboration],
  Phys.\ Rev.\ D {\bf 90}, 052004 (2014)  [arXiv:1406.3827 [hep-ex]];
  V.~Khachatryan {\it et al.}  [CMS Collaboration],  
  Eur.\ Phys.\ J.\ C {\bf 75}, no. 5, 212 (2015)  [arXiv:1412.8662 [hep-ex]].

\end{thebibliography}
\end{document}